\documentclass[preprint]{aastex}
\everymath{\rm}
\everydisplay{\sf}
\received{}
\revised{}
\accepted{}
\cpright{}{}\journalid{}{}
\articleid{}{}
\paperid{}
\slugcomment{To appear in {\it The Astronomical Journal}}
\shortauthors{Kwitter \& Henry}
\shorttitle{Testing S, Cl, \& Ar Nucleosynthesis in PNe}
\begin{document}
\title{Sulfur, Chlorine, and Argon Abundances in Planetary
Nebulae. IV: Synthesis and the Sulfur Anomaly}

\author {R.B.C. Henry}

\affil{Department of Physics \& Astronomy, University of Oklahoma,
Norman, OK 73019; henry@mail.nhn.ou.edu}

\author{K.B. Kwitter}

\affil{Department of Astronomy, Williams College, Williamstown, MA
01267; kkwitter@williams.edu}

\and

\author{Bruce Balick}

\affil{Department of Astronomy, University of Washington, Box 351580, Seattle, WA 98195; balick@astro.washington.edu}

\begin{abstract}

We have compiled a large sample of O, Ne, S, Cl, and Ar abundances which have been determined for 85 galactic planetary nebulae in a consistent and homogeneous manner using spectra extending from 3600-9600~{\AA}. Sulfur abundances have been computed using the near IR lines of [S~III] $\lambda\lambda$9069,9532 along with [S~III] temperatures. We find average values, expressed logarithmically with a standard deviation, of log(S/O)=-1.91$\pm$.24, log(Cl/O)=-3.52$\pm$.16, and log(Ar/O)=-2.29$\pm$.18, numbers consistent with previous studies of both planetary nebulae and H~II regions. We also find a strong correlation between [O~III] and [S~III] temperatures among planetary nebulae. In analyzing abundances of Ne, S, Cl, and Ar with respect to O, we find a tight correlation for Ne-O, and loose correlations for Cl-O and Ar-O. All three trends appear to be colinear with observed correlations for H~II regions. S and O also show a correlation but there is a definite offset from the behavior exhibited by H~II regions and stars. We suggest that this S anomaly is most easily explained by the existence of S$^{+3}$, whose abundance must be inferred indirectly when only optical spectra are available, in amounts in excess of what is predicted by model-derived ionization correction factors. Finally for the disk PNe, abundances of O, Ne, S, Cl, and Ar all show gradients when plotted against galactocentric distance. The slopes are statistically indistinguishable from one another, a result which is consistent with the notion that the cosmic abundances of these elements evolve in lockstep.

\end{abstract}

\keywords{ISM: abundances -- planetary nebulae: general -- 
stars: evolution}

\clearpage

\section{INTRODUCTION}

The abundances and chemical histories of the elements sulfur, chlorine, and argon are important to study because this information provides valuable constraints on stellar evolution theory including yield predictions for massive stars. These three elements, the most abundant isotopes of which are $^{32}$S, $^{35,37}$Cl, and $^{36}$Ar, are produced during both hydrostatic and explosive oxygen burning (Pagel 1997; Woosley \& Weaver 1995). In addition, they all produce prominent emission lines in gaseous nebulae, and so their abundances, particularly those of S and Ar, have been measured in numerous H~II regions, supernova remnants, and planetary nebulae (PNe) in the Milky Way as well as in other galaxies. For objects located within the disk of a single galaxy abundance measurements can be combined with galactocentric distances and kinematics to provide valuable probes of galactic chemical evolution.

Planetary nebulae ostensibly serve as excellent probes of the interstellar abundances of S, Cl, and Ar at the time that their progenitor stars formed, generally more than a billion years in the past. PNe are produced when intermediate mass stars (IMS) with birth masses between 0.8 and 8~M$_{\odot}$ shed a portion of their atmospheres late in the AGB stage of their evolution. This material is thought to have been enriched during prior episodes of dredge-up with products of core and shell H and He burning such as He, C, and N, consistent with the results of direct abundance measurements. Clearly, then, abundances of these three elements in PNe, while crucial for studying IMS nucleosynthesis, are not good gauges of the interstellar levels of these elements at the time of star formation.  Likewise, it is also possible, though unconfirmed, that O in PNe is not a reliable gauge of its interstellar level either, since it may be enriched in PNe through the dredging up of He burning products containing large amounts of $^{16}$O or depleted through ON cycling during H burning. In addition, the Ne abundance may be enhanced in PNe through the conversion of $^{14}$N to $^{22}$Ne. So, the PN abundances of O and Ne, in addition to those of He, C, and N may prove to be unreliable measures of interstellar levels at the time the progenitor stars were formed.  
At the same time, the abundances of S, Cl, and Ar are presumed to remain unchanged from their levels in the ISM at the time and location of star formation; there is currently no theoretical prediction to the contrary. 

Early PN abundance work on S, Cl, and Ar includes the optical studies by Barker (1978a), in which he measured S abundances in 37 galactic PNe using spectroscopic measurements which included the important nebular lines of [S~III] $\lambda\lambda$9069,9532 along with [S~II] $\lambda\lambda$6716,6731. This paper was followed up by a closer study of 20 of those same objects (Barker 1978b) and a third paper on three halo PNe (BB-1, K648, and H4-1) (Barker 1983), using similar techniques. Barker (1980) also reported on observations of [Ar~III] $\lambda$7135 in the same three halo PNe. In the end, he concluded that the S abundance in most PNe is roughly solar but that in the halo objects both S and Ar abundances are markedly lower, in qualitative agreement with Hawley \& Miller's (1978) S measurement in H4-1 and later confirmed by Barker \& Cudworth (1984) in DdDm-1. Other confirmations of low S and Ar in halo PNe were supplied by Torres-Peimbert \& Peimbert (1979), and Pe{\~n}a, Torres-Peimbert, \& Ruiz (1991). Since Barker also found the O and Ne abundances to be both less depleted with respect to solar and uncorrelated with S and Ar, he suggested that the former two elements may be enhanced by nuclear reactions in the PN progenitors through the nuclear processes discussed above, making S and Ar perhaps a better gauge of progenitor composition. Two large optical studies by Aller \& Czyzak (1983) and Aller \& Keyes (1987) of 41 and 51 galactic PNe, respectively, provided S, Cl, and Ar abundances for many more objects with the suggestion that on the average these three elements tend to have subsolar abundances.

While optical spectra permit direct observation of S$^+$ and S$^{+2}$ through the measurement of [S~II] $\lambda\lambda$6716,6731 and either [S~III] $\lambda$6312 or the two near IR (hereafter NIR) [S~III] lines at $\lambda\lambda$9069,9532, photoionization models suggest that S$^{+3}$, which has a strong emission line in the infrared but none in the optical, may also be an abundant ion in PNe. In work limited to the optical, then, the abundance of S$^{+3}$ is customarily accounted for through the use of an ionization correction factor (ICF). The sulfur ICF is discussed in detail in Kwitter \& Henry (2001; Paper~I) as well as in {\S}4.2.5 below. 

Observation of the S$^{+3}$ emission line at 10.5$\mu$m in the infrared allows direct measurement of the S$^{+3}$ abundance. This abundance, when added to the optically observed S$^+$ and S$^{+2}$ abundances\footnote{Care must be taken to ensure that the nebular regions observed and the aperture sizes utilized in the optical and IR are the same, or that differences are accounted for as fully as possible.}, gives total S, eliminating the need for an ICF. Such an approach was pioneered by Dinerstein in her thesis (Dinerstein 1980a) and in Dinerstein (1980b), in which she measured the [S~IV] 10.5$\mu$m emission in 12 galactic PNe. Combining these measurements with optical data obtained by Lester, Dinerstein, \& Rank (1979), abundances of the three S ions were computed and S/H ratios established. Their results indicated that S/H in their sample objects was roughly solar. Then Garnett \& Lacy (1993) employed similar IR techniques to study S in two halo PNe, K648 and BB-1, and were able to reconfirm Barker's low S/O values in these objects. Very recently Dinerstein et al. (2003) have extended their work to include IR observations of two halo PNe, DdDm-1 and H4-1. They also find low S/O ratios and suggest, as Barker and Garnett \& Lacy had, that O enrichment had occurred in these low metallicity objects, an idea also proposed by P{\'e}quignot et al. (2000) following the analysis of two PNe of roughly half-solar metallicity in the Sagittarius dwarf galaxy. Finally, IR observations were also used by Beck et al. (1981) to infer S and Ar abundances for 18 galactic PNe. This group found some deviation from the expected lockstep behavior of S and Ar.

Since the late 1980s numerous additional optical studies of S, Cl, and Ar in PNe have appeared. These include a survey of 14 Galactic objects by Guti{\'e}rrez-Moreno \& Moreno (1988), 43 Galactic objects by Freitas Pacheco et al. (1991; 1992), 15 Galactic PNe by Costa et al. (1996), 23 LMC PNe by Freitas Pacheco et al. (1993a,b), 80 Galactic objects by Kingsburgh \& Barlow (1994; hereafter KB), and 15 PNe (twelve Galactic and three in the LMC) by Tsamis et al. (2003). Howard, Henry, \& McCartney (1997) carried out a detailed photoionization analysis of nine halo PNe and included S and Ar in their list of elements. Finally, Maciel \& K{\"o}ppen (1994), Maciel \& Chiappini (1994), and Maciel \& Quireza (1999) published large compilations of S, Cl, and Ar abundances and applied the data to the study of galactic chemical gradients.
The general conclusion that can be gleaned from the above studies is that S/O is often found to be less than either the solar or H~II region value, while Cl/O and Ar/O appear to be consistent with these levels\footnote{Here, the solar value for 12+log(O/H) has been taken to be 8.69, as published by Allende Prieto, Lambert, \& Asplund (2001).}.

Despite the large amount of S, Cl, and Ar abundance information already available for PNe, we undertook a new study of these elements five years ago, because we were aware that most previous S studies had been based on the [S~III] auroral line at 6312~{\AA}, although potentially better abundance information could be derived using the strong lines of [S~III] $\lambda\lambda$9069,9532. In addition, we were intrigued by the implication, already noticed by Freitas~Pacheco (1993), that significant amounts of S, Cl, and Ar may be produced by Type~Ia supernovae (Nomoto et al. 1997), and thus we wanted to see if, in the possession of consistently determined data for a large PN sample, along with published stellar yields, we could test the hypothesis that SNIa make noticeable contributions to the cosmic buildup of these three elements.

The current paper is the fifth one in a series dealing with an extended project in which the abundances of S, Cl, and Ar (in addition to He, N, O, and Ne) have been determined for 85 galactic PNe. 
In the previous four papers, spectral data and abundance calculations were presented for subsets of this sample: Kwitter \& Henry (2001; Paper~I), Milingo, Kwitter, \& Henry (2002; Paper~IIa), Milingo, Henry, \& Kwitter (2002; Paper~IIb), and Kwitter, Henry, \& Milingo (2003; Paper~III). The present paper attempts to collate all of the abundances previously reported and to compare the large sample with analogous information from other PN samples as well as H~II regions in order to identify and comment upon trends. In section~2, we briefly discuss the nature of our PN sample, while in section~3 we present results of our study pertaining to electron temperatures. Section~4 contains an extensive discussion of our abundance results, while we present our results pertaining to Galactic chemical gradients in section~5. Our conclusions are presented in section~6. Future papers will address questions related to population, morphological type, and abundances as well as the potential impact of Type~Ia supernovae on the cosmic buildup of S, Cl, and Ar.

\section{THE DATA}

The sample of PNe which is the subject of this paper comprises 85 Galactic 
objects\footnote{Note that NGC~2242 was part of the original sample, but we have eliminated it from consideration in this paper, as its extremely high excitation level makes the derived abundances uncertain (Paper~III).}. Table~1 contains the common name of each object in the first column 
along with the Peimbert type in the second 
column. By type our sample includes 12 Type~Is, 69 Type~IIs, and 4 halo or 
Type~IVs. These types, originated by Peimbert (1978), classify PNe
primarily according to chemical composition, a proxy for their progenitors' 
galactic population characteristics. Type~Is are those with enhanced 
nitrogen (and often, helium) abundances, assumed to result from 
younger, more massive progenitors; Type~IIs are intermediate population with 
less enhancement; Type~IIIs have similar abundances to Type~IIs, but in 
addition have higher peculiar velocities; and Type~IVs are halo PNe. 

The original Type~I classification included both enhanced He/H and N/O. In 
this work, we follow the discussion found in KB, 
who use only the nitrogen abundance as a discriminant. They set the minimum nitrogen 
abundance for Type~I classification in a given galaxy equal to the sum of the carbon 
plus nitrogen abundances of H~II regions in that galaxy. Assignment of 
Type~I status thus requires that nitrogen must have been produced by 
conversion of carbon in the envelope of the progenitor star (so-called 
"hot-bottom burning"). KB calculated the equivalent 
minimum N/O ratio for Type~I PNe in the Galaxy to be 0.8. Using more recent abundance 
measurements for the Orion Nebula (Esteban et al. 1998) and for the sun 
(Grevesse \& Sauval 1998), we derive a somewhat lower minimum N/O ratio of 
0.65, which we have applied here. 

Our sample was purposely chosen to contain a large majority of Type~II 
objects. These objects are ideally suited for probing the interstellar 
medium, as they are known to be disk objects with relatively lower 
progenitor masses than the Type~Is, and hence they are less likely to be 
self-contaminated with nucleosythetic products. 
An important selection goal was to produce a program list 
which collectively spanned a large range in galactocentric distance. In this 
respect, our objects range from 2-17~kpc. 

Spectrophotometric data covering the region between 3600 and 9600~{\AA} were 
obtained for each of our program objects using either the 2.1~m telescope and 
Goldcam spectrograph at KPNO or the 1.5~m telescope and the Cassegrain 
spectrograph at CTIO between 1996 May and 1999 July. The data were reduced, 
dereddened, and measured using standard IRAF routines. Abundances were uniformly
calculated using the program and procedure discussed in detail in Paper~I. 
Specific references for the atomic data used in the abundance calculations 
are listed in Table~4 of that paper along with an extended discussion of the 
ionization correction factors employed. 

Above all, our goal has been to produce a large homogeneous sample of PN 
abundances starting with state of the art optical spectrophotometry and 
using up-to-date atomic data consistently throughout our study as much as 
possible. Researchers can view and manipulate the reduced spectra for each 
of our objects at our {\it Gallery of Planetary Nebula Spectra} 
website\footnote{http://cf.williams.edu/public/nebulae/index.html}. 
Additional information as well as links to images of each object and a set of student lab exercises using these data will 
be found there. Questions regarding the website or the original data or should be addressed to K.B. Kwitter 
(kkwitter@williams.edu). 

\section{ELECTRON TEMPERATURES}

Electron temperatures and densities for our sample objects are compiled in Table~2. The references for these values are identical to those listed in the last column of Table~1. Due to an oversight in earlier papers, effects of density were not accounted for in the determination of the [N~II] temperature in those objects where the density exceeds 5000~cm$^{-3}$. This problem has been addressed, and the updated values for the relevant objects now appear in Table~2. 
Fig.~1 shows five electron temperatures, [O~III], [N~II], [O~II], [S~II], and [S~III] plotted together in various combinations for our objects, where temperatures are in units of 10$^4$K, and the diagonal lines show one-to-one relations\footnote{The specific lines used for temperature determinations are as follows. [O~III]: nebular, $\lambda\lambda$4959,5007, auroral, $\lambda$4363; [N~II]: nebular, $\lambda\lambda$6548,6584, auroral, $\lambda$5755; [S~III]: nebular, $\lambda\lambda$9069,9532, auroral, $\lambda$6312; [O~II]: nebular, $\lambda\lambda$3726,3729, auroral, $\lambda$7325 (quartet); [S~II], nebular, $\lambda\lambda$6716,6731, auroral, $\lambda$4072 (quartet).}. Each panel is labeled to show the specific temperatures being plotted, in Y v. X format, along with a representative error bar of $\pm$1000K. 

In the left panels of Fig.~1 there are suggestions of correlations between the [S~II] and [O~II] temperatures (top left), and between each of these and the [N~II] temperature (middle and lower left), although the scatter is large in each case. The ionization structure of model nebulae indicates that these three ions should coexist spatially, and so it is somewhat surprising that the correlations in the left panels involving these temperatures are not greater. In the right panels we have plotted the [N~II], [O~II] and [S~III] temperatures as a function of the [O~III] temperature. In the upper panel, there is a suggestion of a positive correlation between the [N~II] and [O~III] temperatures below 10,000~K for both temperatures, but above that level we see a large amount of scatter. 

The most interesting result in Fig.~1 is the relation between the [O~III] and [S~III] temperatures shown in the lower right panel, a relation previously seen in H~II region studies by Garnett (1992), Vermeij \& van~der~Hulst (2002), and Kennicutt, Bresolin, \& Garnett (2003; hereafter KBG). Systematically, we find [S~III] temperatures are greater than [O~III] temperatures in PNe, and the difference seems to grow with [O~III] temperature. A linear fit to the trend is shown in the figure as a solid bold line with a correlation coefficient of 0.79 and an equation of the form 
\begin{equation}
T_{[S~III]}=-0.039(\pm.11)+1.20(\pm0.11) \times T_{[O~III]},
\end{equation}
where both temperatures are in units of 10$^4$~K. We also show for comparison purposes the analogous form derived by Garnett (1992), $T_{[S~III]}=0.17 + 0.82T_{[O~III]}$, which he derived from a photoionization model study of H~II regions. Clearly,  for PNe the latter expression underestimates the [S~III] temperature for a given [O~III] temperature, but this may be related to general density differences between PNe and H~II regions, since the greater values seen in the former may result in a different thermal structure in these objects.
Our expression can be used to obtain [S~III] temperatures from [O~III] temperatures with an uncertainty of $\sim$1000K. Researchers may find this relation particularly useful for obtaining S$^{+2}$ abundances when the lack of [S~III] nebular lines due to spectral limitations prohibits the direct determination of the [S~III] temperature.

There is noticeable scatter in all of the panels of Fig.~1. In some cases, the cause could be that the two relevant ions do not exist in the same region of the nebula. For example, photoionization models suggest that
O$^+$ and N$^+$ do not coexist spatially with O$^{+2}$. Thus we might expect scatter when their temperatures are plotted against the [O~III] temperature. Additional factors which lead to scatter in [O~II] temperatures were explored by KBG and include radiative transfer effects and shock heating. Additional sources of scatter may be uncertainties in line intensities and reddening. Finally, variations in density and temperature along the line of sight could introduce scatter.  In the low density limit, observed line intensities are line-of-sight integrations of the product of density and excitation rate, where the latter has a maximum value at a specific optimum temperature which generally is not the same for both the nebular and auroral lines comprising the temperature diagnostic. This means that the regions of the nebula responsible for the bulk of the emission may be different for the two line types, and thus may be weighted by different densities. In this way, an observed nebular to auroral line strength ratio may not imply the true representative temperature, i.e. the presence of such inhomogeneities may cause the line ratios used in determining temperatures to be skewed up or down.

\section{ABUNDANCES}

\subsection{General Results}

A complete summary of abundances for He/H, O/H, Ne/O, S/O, Cl/O, Ar/O, and N/O\footnote{The N/O ratios for those objects for which temperatures have been corrected for density effects (see {\S}3) have likewise been updated and included in this table.} for our 85 survey objects is provided in Table~1. The first column gives the object name and the second column the Peimbert type according to the criterion described in {\S}2 (N/O$\ge$0.65). We also provide an estimate of the galactocentric distance of each object in kiloparsecs, where the distances were determined using basic information in the references cited in the relevant table footnote. The abundances were derived from spectra reported and analyzed in the preceding papers in this series; information in the last column of the table gives the reference source for the abundances of a particular object. Also, we have included 10 objects in Table~1 for which line strengths and some abundance results were reported in an earlier series of papers. We recalculated all abundances for these 10 objects, including those of S, Cl, and Ar which were excluded in the original determinations, and include them in Table~1. A tabulation of ionic abundances, electron densities and temperatures for these objects is provided in Appendix~A. Finally, the last four rows of Table~1 give averages for sample subsets as well as for the entire group of objects. Uncertainties in elemental abundances given in the table are ultimately derived from line strength uncertainties and their effects on ionic abundances. Uncertainties in the latter were added in quadrature to produce the elemental abundance final uncertainty. Readers are referred to the earlier papers in this series for a detailed discussion of our abundance-determining methods.

Table 3 provides a comparison of our abundance ratio averages for S/O, Cl/O, and Ar/O with those found in numerous other studies of PNe and H~II regions, as well as values observed in the Sun (Grevesse \& Sauval 1998). Column~1 gives the source reference, while column~5 provides comments pertaining to each sample. Values given in columns~3, 4, and 5 are logarithms of arithmetic averages. 
In the case of PNe, the number of sample objects contributing to an average value is given in parentheses. It should be pointed out that our PN study is the only one on the list that utilized the NIR lines of [S~III]; in all of the other cases S$^{+2}$ abundance computations were based on the $\lambda$6312 line. 

The three ratios considered in Table~3 are of interest in that they involve elements whose abundances ostensibly vary in lockstep, according to standard nucleosynthesis theory. Each ratio, then, should provide a key check on the relative stellar yields for the two elements, and they should show consistency between PNe and H~II regions, since their values in the former should not be altered by nuclear processes in the progenitor star but reflect interstellar values at the time of star formation. Indeed the average abundance ratios for Cl/O, and Ar/O are very consistent between all of the PNe and H~II region samples as well as in the Sun, in accord with the above ideas. {\it However, based upon the numerous PN studies included in Table~3, S/O seems to be less in PNe than in H~II regions and the Sun.} Our results are no exception.

The four panels in Fig.~2 show separately the correlations between Ne, S, Cl, and Ar, versus O for both our sample objects (open circles) and H~II regions (filled symbols). The bold, solid line in each panel shows a least squares linear fit to the data, where the values for slope, intercept, correlation coefficient and number of included sample objects are listed in Table~4. In three panels the H~II region abundances reported by KBG for M101 are shown with filled circles. We include their data for comparison, because it is characterized by high signal-to-noise and, in the case of sulfur, they derive S$^{+2}$ abundances using the NIR lines, as we did.

We note in Table~4 that the correlation coefficient for Ne is especially high, and the relation between Ne and O found here is remarkably similar to the one found by Henry (1989) in which the intercept was -2.14($\pm$.04), the slope was 1.16($\pm$.33), and the correlation coefficient was 0.91. The current data simply reconfirm the tight relation between these two elements. In addition, we see that the M101 H~II region data fall very nicely along the linear fit for the PNe, in strong support of the ideas that these elements are forged together, vary in lockstep, and that PN measurements of these elements measure interstellar values, with no discernable contamination from or depletion by the progenitor star.

In the cases of Cl and Ar, however, the scatter in Fig.~2 is observed to be slightly greater than for Ne, especially for PNe. Generally speaking, these elements are products of both hydrostatic and explosive oxygen burning and require several alpha reactions to produce them, perhaps making their production rates more vulnerable to local conditions. In addition, the line strengths required to measure their abundances are generally weaker than those for O, Ne, and S. Thus, the larger scatter is anticipated. Note also that the M101 Ar abundances appear to be slightly larger systematically compared to the PN Ar abundances.

We have identified several outliers in three of the panels in Fig.~2. It is interesting to note that three of these objects, BB1, H4-1, and K648, are halo PNe. Howard et al. (1997) found that abundances in halo PNe often exhibited inconsistent behavior with respect to disk objects, a pattern also seen in abundance studies of halo stars (Sneden \& Cowan 2003; Truran et al. 2002).

{\it But by far the most intriguing result in Fig.~2 is that PNe behave differently from the H~II regions of M101 in the S-O plane.} In the Introduction, we briefly discussed a few cases of individual low-metallicity PNe in which the S/O ratio was found to be subsolar. Specifically, papers by Barker (1983), Howard, Henry, \& McCartney (1997), P{\'e}quignot et al. (2000), and Dinerstein et al. (2003) make note of this anomaly. Yet here we find the problem to be pandemic, stretching broadly across the metallicity range for a large sample of objects. Recall that in our discussion of Table~3 we pointed out that average S/O values in PN samples tend to be less than in H~II regions. In Fig.~2 we see just how extensive this pattern is. Interestingly, this general trend has not been noticed before, perhaps because a broad, detailed intercomparison among PNe and H~II regions and the Sun for these elements has not previously been made.

Assuming that H~II region S and O abundances represent the true interstellar values\footnote{Chemical evolution theory predicts that a unique, spatially independent relation exists between oxygen and sulfur (and other alpha elements) production over the entire metallicity range such that the two elements evolve in lockstep. This situation arises because both elements are produced predominantly in massive stars. The observed correlation between S/H and O/H abundances in H~II regions residing in various galaxies with different rates of evolution and age-metallicity relations ({\S}4.2) strongly supports this idea observationally. For our PN observations of low S/O to be an effect of chemical evolution would seem to require that the interstellar gas which formed these PNe was processed under vastly different conditions than was the gas which formed the H~II regions. e.g. a much different stellar initial mass function. For the time being we rule out this possibility.}, then is the depressed S/O value in PNe relative to H~II regions produced by low S abundances or high O abundances? The suggestion has been made in a few of the papers cited above that the anomalously low S/O in a few metal-poor PNe might be the result of O synthesis as the result of He burning in the progenitor star. However, the positions of PNe in the Ne-O plane seem to argue otherwise. For example, the PNe and H~II regions fall along the same line in the Ne-O plane (upper left panel), suggesting that if O enrichment is indeed occurring as proposed, then so too is Ne enrichment and in the same proportion, since otherwise, the PNe would be offset to the right of the H~II region track. But since it is unlikely, due to temperature constraints, that excess $^{20}$Ne is produced by the same He burning that allegedly produces the O, then excess Ne would have to come from $^{22}$Ne production from $^{14}$N, and it is difficult to imagine that this channel would operate at just the proper rate necessary to maintain the PN positions on the Ne-O track in Fig.~2. Finally, if the S-O anomaly in PNe is the result of problems inherent in the abundance determinations, again the tight Ne-O correlation for PNe suggests that the culprit must be S, not O. Therefore, we conclude that the S-O anomaly in the upper right panel is most likely related to S abundances, not O abundances, and is therefore a S anomaly. In the following subsection we proceed with that assumption in an effort to understand the origin of the anomaly.

\subsection{The Sulfur Abundance Anomaly}

To study the S abundance anomaly in more detail, we add other H~II region
samples, as well as stellar samples, to the S-O plot of Fig.~2 and display
the result by itself in enlarged format in Fig.~3a.
Noting that the combined H~II region and stellar data form a rather narrow,
linear track extending over roughly 1.5~dex along both axes, we make the
following points: 1)~Compared with H~II regions
and stars, PN sulfur abundances are systematically lower, particularly at
subsolar oxygen levels; and 2)~PNe exhibit a large amount of
scatter in the S-O plane, while H~II regions show a relatively tight
correlation. To make the point that
the S anomaly is investigator-independent, we plot KB's abundances for their
large sample of PNe, derived using the $\lambda$6312 line, in Fig.~3b along
with the same H~II region-stellar comparison data shown in Fig.~3a. KB's
abundances are taken from their Table~13. Clearly, the anomaly is not new,
just heretofore unrecognized. Assuming that the
track defined by the H~II region data in Fig.~3a defines the true
interstellar relation between S and O, the sulfur anomaly refers to the situation in which many PNe appear to have a
deficient amount of S relative to the interstellar value associated with their O/H ratio. We now briefly explore possible explanations
for the sulfur anomaly.

If our sample of PNe are drawn from a
population which is distinct from H~II regions, then the abundance of one element
with respect to another can be much different, perhaps due to mixing of
star-forming gas. Witness the numerous examples common in the literature
today in which abundance patterns in metal poor systems exhibit considerably
more scatter and incoherence than do patterns associated with disk systems
(McWilliam 1997). A detailed study of our sample with regard to population
is beyond the scope of this paper. However, simple comparisons of S/O ratios
of our PNe with their galactocentric distances and heights above the plane
currently reveal no systematic behavior. At the same time, we expect that
our sample of objects is relatively homogeneous, in that we purposely chose
a group of objects the vast majority of which are Peimbert Type~II PNe,
meaning that they currently are part of the disk population. So, we do not
expect population-related factors to be the cause of the S anomaly.

Another possibility is that sulfur is depleted onto dust in many PNe.
However, Savage \& Sembach (1996) found that sulfur is not greatly
refractory and does not form dust readily, so this explanation for 
the anomaly seems unlikely. Likewise, nucleosynthesis arguments appear untenable. In Fig.~2 we saw that Ne, Cl, and Ar in both H~II regions and PNe follow a near lockstep behavior with O, and it seems likely from a theoretical point of view that S would behave in a similar way.

The most likely cause of the S anomaly, then, is some factor related to the determination of the S abundance itself.
In an effort to spot potential fundamental problems in our calculations, we compared our S abundances with those of KB for the PNe appearing in both samples and found satisfactory agreement between the two sets of values. Along the same lines, we used our methods to calculate S abundances using KBG's H~II region data and found very close agreement with their published abundances. Finally, we recalculated all of our S$^{+2}$ abundances using the [S~III] $\lambda$6312 line rather than the NIR lines. This method produced abundances which were slightly higher systematically\footnote{This was most likely because [N~II] temperatures were used in place of the generally higher [S~III] temperatures.}, although the offset was entirely insufficient for explaining the sulfur anomaly.

Next, in an attempt to independently confirm the S anomaly using a different abundance technique, we calculated photoionization models for IC~4593, Hu2-1, and NGC~3242, three
PNe whose S abundances fall significally below the S-O relation defined by
H~II regions.
We chose these three particular objects because their C
abundances have been measured (Henry, Kwitter, \& Bates 2000), an important
factor, since C can be a significant coolant and thus a parameter that needs
to be constrained if possible. We used the photoionization code CLOUDY
version 90.4 (Ferland 1996). Important observationally constrained input
parameters were the central star temperature, and the nebular H$\beta$
luminosity, density, radius, and chemical composition. We proceeded by
setting the model parameters, including abundances, equal to the observed
ones, calculating a model, comparing output line strengths with the
observations, making changes to the input parameters, and continuing to
iterate until model and observed line strengths matched closely. We assumed
a blackbody shape for the stellar spectrum along with a constant density
spherical nebula. 

Table~5 lists the observed and model-predicted line strengths by object in
the upper part of the table, while in the lower section the observed and
model-input parameter values are provided. Observed quantities were taken
from sources detailed in the table footnotes. The first eight lines of model
parameters give abundance information, followed by the central star's
effective temperature, the nebular H$\beta$ luminosity, electron density,
radius, and filling factor.

The two important points to recognize in Table~5 are that the model output line
strengths agree closely with the observations for all three objects, and that
the observed abundances, i.e. those in Table~1, agree well with the ones necessary in the model to produce the line strength
agreement. In particular the model O and S abundances agree with their
observed counterparts to within about 0.1~dex. To emphasize this point, the
positions of the models in S-O abundance space are shown in Fig.~3a with
large stars and labelled by object name. Note that the model and observed
positions for IC~4593 and NGC~3242 are offset slightly (the two lines point
to the model and the observed data point) while for Hu2-1 the model and
observed points are nearly on top of one another. We conclude that photoionization models, which are heavily constrained by
numerous observed quantities, imply S and O
abundances which agree closely with those already inferred and confirm the existence of the S anomaly in these three objects. Therefore,
if there is a problem with our analysis procedure,
then the trouble extends to the models as well. Such a problem is likely to be related to our lack of understanding of the details of the ionization structure of objects exhibiting the S anomaly.
{\it We propose that the probable cause of the S anomaly is the failure to adequately measure indirectly, through the use of an ICF, all of the S$^{+3}$ present in those nebulae which display a large S deficit.} In other words, common forms of the sulfur ICF must be flawed.

Any choice of ICF for sulfur can, in principle, be checked
empirically by comparing the S$^{+3}$ abundances it implies\footnote{S$^{+3}$=(S$^+$+S$^{+2}$) $\times$ (ICF-1)} with directly
observed values in objects where the [S~IV] 10.5$\mu$m strength has been measured. Table~6 provides such a comparison for 14 of our
objects for which IR data are available. In columns~2 and 3 we give the
S$^{+3}$/H$^+$ abundance ratio as inferred from our original ICF and from
the IR data, respectively, while column~4 gives the ratio of these two quantities. The first number in the last column indicates the source for the column~2 abundances, while the second number does the same for column~3. For reference~6, the ratios quoted here were actually
computed using our ionic abundance routines along with Dinerstein's (1980b) original line strengths for [S~IV]. We employed the [S~IV] collision strengths of Tayal (2000). Numbers in column~4 indicate a general tendency for our ICF method to underestimate the S$^{+3}$ contribution to the S abundance, a trend that can be seen graphically in Fig.~4, where we plot S$^{+3}$ abundances
derived from the IR against the same parameter inferred from the ICF. In Table~6 the average of the ICF-to-IR values between 0 and 1 is .55
or -.26~dex. Interestingly, the latter offset is consistent with the amount by which objects in Fig.~3a tend to fall below the H~II regions, consistent with our contention that our ICF undercorrects for S$^{+3}$. (As a caveat, we point out that positions and aperture sizes employed for the optical and IR observations of each PN do not necessarily coincide for the 14 objects analyzed here, so the results of this comparison are only tentative. To improve on this analysis idea, therefore, it will be necessary to sample many more PNe, using
consistent aperture sizes and positioning.)

Finally, we define the sulfur deficit as the magnitude of the vertical offset of a PN from the H~II region track at a specific metallicity (O/H) in the S-O plane and assume that it represents the amount of S$^{+3}$ which is unaccounted for through our methods. We have estimated the sulfur deficit for each
object in our sample by employing a least squares fit to the H~II region data in Fig.~3a\footnote{$12+log(S/H)_{H~II} = -3.37 + 1.22[12+log(O/H)]$}. Upon plotting the deficit values against various other quantities, an
interesting correlation with O$^{+2}$/O abundance ratios was found. This relation is displayed in Fig.~5, where the sulfur deficit is plotted on the vertical axis in units of 10$^5$S$^{+3}$/H$^+$. Note that PNe with greater sulfur deficits are positioned higher in the plot. There is a
slight tendency for the S deficit to be higher in
objects where the O$^{+2}$/O ratio is low. The interpretation of this, however, is unclear, since O$^{+2}$/O can be small either because of high or low excitation, and likewise because of matter or radiation boundedness.
But whatever the explantion, more S$^{+3}$ than expected tends to be present when this ratio is low. 
The sulfur deficit could arise if a process contributing to the
ionization balance between S$^{+3}$ and S$^{+2}$ is unaccounted for.
Candidates include a higher than expected S$^{+2}$ photoionization
cross-section, a low rate of recombination for S$^{+3}$, peculiarities in
the central star spectrum which selectively force the ionization equilibrium
upwards, or highly matter-bounded gas dominated by S$^{+3}$. We emphasize
that these possibilities need not be inconsistent with the fact that H~II
regions do not exhibit the S anomaly, since in the latter objects the
excitation level is much lower and the abundance of S$^{+3}$ is expected to
be extremely low. Any of the above possibilities would not likely be evident
in the observed ion abundances.

\section{GALACTIC ABUNDANCE GRADIENTS}

Figures 6a-e show the abundances of O, Ne, S, Cl, and Ar, expressed as 12+log(X/H), as functions of galactocentric distance in kiloparsecs in the galactic disk as tracked by our PNe. Our data, taken from Table~1, are shown as open circles in each case. Fig. 6a shows our results for oxygen along with the H~II region and stellar data taken from the literature and detailed in the figure captions. Figures 6b-e give analogous results for Ne, S, Cl, and Ar, respectively. Data for the halo PNe K648, DdDm-1, BB1, and H4-1 are not included in these plots. The solar value indicated in each plot by a large S is the photospheric value of Allende Prieto et al. (2001) in the case of oxygen, while solar abundances for the other elements were taken from Grevesse \& Sauval (1998): Ne (photospheric), S (meteoritic), Cl (meteoritic), and Ar (photospheric). A representative error bar relevant to our data is given in each plot. Finally, we remind the reader that typically PN distances contain relatively large uncertainties, and thus some of the scatter in each of the figures is undoubtedly the result of this fact.

Table~7 gives the parameters and information related to linear fits to our results for the five elements in Figs.~6a-e. For each element indicated in the first column we give the y-intercept and slope, along with a correlation coefficient, number of sample objects included in the fitting, and the logarithmic offset from the solar value at the distance of 8.5~kpc, where the solar values are taken from the references cited in the previous paragraph.

Of particular interest are the slopes in column~3, since comparing values for Ne, S, Cl, and Ar with O may either confirm or argue against the often-assumed lockstep chemical evolution of the first four elements with the last one. We see that considering the uncertainties, all of the slopes agree with one another, and so we find no evidence in our data that such lockstep behavior does not take place. This is in qualitative agreement with Garnett's (1989) conclusion following his detailed study of S in extragalactic H~II regions.

For oxygen, Fig.~6a shows the consistency which exists among data from both nebular and stellar sources located in the disk. Our gradient of -0.037$\pm$.008 dex kpc$^{-1}$ agrees within the uncertainties with the value derived by Deharveng et al., i.e. -0.0395$\pm$.0049 dex kpc$^{-1}$; like them we see no suggestion that the gradient flattens in the outer regions. Our measurement is somewhat less than values given by Afflerbach et al., -0.064$\pm$.009 dex kpc$^{-1}$, Gummersbach et al., -0.07$\pm$.02 dex kpc$^{-1}$, and Rolleston et al., -0.067$\pm$.008 dex kpc$^{-1}$. Deharveng et al. obtain a correlation coefficient of -0.86, which is signficantly better than our value of -0.45. It is very likely that the larger uncertainties inherent in PN distances explain much of this difference. We note, for example, that the coefficients reported by Maciel \& Quireza (1999) for gradients derived from PNe are generally less than H~II region values for the same element. Nevertheless, using our gradient, our interpolated value for 12+log(O/H) at the solar circle (8.5~kpc) is 8.66, close to the solar measurement by Allende Prieto et al. (2001) of 8.69 as well as the one by Holweger (2001) of 8.74. We mention that while distances to B stars are likely to be much better known than PN distances, these objects nevertheless exhibit some scatter in Fig.~6a, suggesting that PN distances are not likely to be the only important source of scatter.

It is interesting to note that inside the solar circle our data are compatible with a relatively flat abundance gradient, with a slight steepening beyond that point. We point out that we do not see evidence for a flatter gradient in the outer disk compared with the inner regions, as reported by V{\'i}lchez and Esteban (1996) and Maciel \& Quireza (1999), although our observations may be consistent with the report by Twarog, Ashman, and Anthony-Twarog (1996) of a discontinuity in [Fe/H] along the Milky Way disk, where [Fe/H] dropped by 0.3 dex beyond 10~kpc. However, we observed only a few PNe beyond 10~kpc. 

Turning to Ne in Fig.~6b, we see that our data are consistent with measurements by Shaver et al. (1983). Our gradient of -0.044$\pm$.014 agrees within the uncertainties with Maciel \& Quireza's measurement of -0.036$\pm$.010. Our interpolated Ne abundance at the Sun's location is 8.00, close to the Grevesse \& Sauval (1998) number of 8.08. Additional comparison values for the solar vicinity are 7.89 for Orion (Esteban et al. 1998) and 8.13 for M17 (Peimbert, Torres-Peimbert, \& Ruiz 1992). However, there is suprisingly little Ne abundance data available especially for H~II regions inside the solar circle; it would be very interesting to probe the Ne gradient in the Milky Way disk to a greater extent than has been done in the past.

The sulfur data displayed in Fig.~6c show the sulfur anomaly (see {\S}4.2) clearly, as our PN data are shifted downward systematically by about 0.5~dex from H~II regions. We see also the relatively larger amount of scatter in the PN abundances than in the H~II regions. Our S gradient of -0.048$\pm$.0098 dex kpc$^{-1}$ is flatter than Maciel \& Quireza's value of -0.077$\pm$.011 dex kpc$^{-1}$, while our interpolated solar S abundance is 6.64 compared with the meteoritic measurement of 7.20 in Grevesse \& Sauval. However, in light of the discussion of the S anomaly above, it is questionable whether PNe can currently be used to probe the galactic S gradient.  Finally, the Cepheid abundances are systematically greater than those for H~II regions by about 0.5~dex.

Fig.~6d shows our results for Cl, where PN data from Maciel \& Chiappini (1994) are included only for comparison purposes, since many of the objects in the two samples are the same. Clearly there is significant scatter in both samples, which is perhaps explained by the relatively weak emission lines associated with Cl. We measure a gradient of -0.045$\pm$.013 dex kpc$^{-1}$, compared with Maciel \& Chiappini's steeper value of -0.07$\pm$.01 dex kpc$^{-1}$. Our interpolated solar value is 5.07, slightly smaller than Grevesse \& Sauval's measurement of 5.28 in the Sun. While Cl abundances in H~II regions over a large range of galactocentric distances are in short supply [Rodr{\'i}guez (1999) reports ionic abundances but no separate elemental abundances for the seven Galactic H~II regions studied], Esteban et al. (1998) find a value of 5.33 in Orion, while Peimbert et al. (1992) infer a level of 5.48 in M17, both somewhat larger than our value for the solar vicinity.

Finally, our Ar measurements are displayed in Fig.~6e. The PN and H~II region data sets are very consistent but with some scatter. We find a gradient of -0.030$\pm$.010 dex kpc$^{-1}$, while Maciel \& Quireza measure a steeper gradient of -0.051$\pm$.010 dex kpc$^{-1}$. Our interpolated solar Ar value is 6.33, while Grevesse \& Sauval's photospheric number is 6.40, close to ours. Additional comparisons are 6.49 for Orion (Esteban et al.) and 6.63 for M17 (Peimbert et al.).

In summary, each of the five elements, O, Ne, S, Cl, and Ar shows a negative gradient with galactocentric distance, as anticipated. The consistency of the slopes among the five elements that we studied adds support to the idea that these elements increase globally in lockstep. The most interesting result continues to be the systematic offset in sulfur abundances between H~II regions and PNe  (Fig.~6c).

\section{CONCLUSIONS}

This paper is the culmination of a project whose object is to study the abundances of S, Cl, and Ar in planetary nebulae in detail. We carefully selected a sample of 86 PNe spanning a large range in galactocentric distance but representing primarily the Galactic disk population. A few halo PNe, however, were included in our sample.

We observed each object spectroscopically between 3600-9600~{\AA} with the express purpose of including the strong nebular NIR lines of [S~III] in our data for each object. Additionally, we have used our data to compute ionic and elemental abundances for these three elements. Our purpose has been to produce a large, homogeneous set of chemical abundances.

In considering the resulting electron temperatures derived from our data for each object, we found only weak correlations between most of them with the exception of the [O~III] and [S~III] temperatures. For this last case, we discuss a close direct relation between these two temperatures, possibly related to the fact that they occupy similar regions of the nebula and that both are associated with lines with high critical density, well above those encountered in PNe.

In compiling our abundance results we found a strong correlation between Ne and O, the robust relation that is consistent with earlier work by ourselves and others. Likewise, Cl and Ar abundances are correlated with O but with much more scatter in the relation. Our average abundance ratios of Cl/O, and Ar/O proved to be very consistent with those derived in other PN studies as well as studies of H~II regions and the Sun. We also found that our average S/O ratio was somewhat below previous findings by other researchers, but still within the scatter and uncertainty of most determinations.

The most interesting correlation is the one between S and O. When compared with similar data for stars and H~II regions, we find that many PNe have much lower S abundances than expected given their O abundances, a result we have called the S anomaly. We emphasize that this is not a new result, as it appears in others' data as well when the same comparisons are made. Our result merely represents the first time that it has been {\it noticed} on such a large scale. We also point out that H~II regions appear not to suffer from this anomaly. In an attempt to discover the anomaly's cause, we have looked carefully at our abundance method for S, comparing with others' results as well as our own detailed photoionization models. We have also employed our ionization correction factor to estimate the amount of unseen S and compared our implied amounts directly with observed ones for the few cases where both types of data are available.

While we were unable to establish with certainty the cause of the S anomaly, we feel it is most likely due to excess amounts of S$^{+3}$ in many nebulae beyond what is implied by models or inferred directly from observations of S$^+$ and S$^{+2}$. Thus, this additional S escapes indirect detection when an ICF is used. Our conclusion is supported by our finding that most direct determinations of S$^{+3}$ using IR spectroscopy imply a larger amount of this ion than is inferred indirectly when only optical spectra are used. This result is uncertain, however, because of inconsistencies in aperture size and nebular position between the observations of the two different spectral regions. In addition, by calculating the amount of S$^{+3}$ necessary to explain the S deficit for each object, we find that there is an inverse correlation between this amount and the O$^{+2}$/O abundance ratio, consistent with the idea that the deficit is greater in low excitation nebulae.

Finally, we analyzed our abundance data in terms of galactocentric distance. We found evidence for abundance gradients for O, Ne, S, Cl, and Ar; due to scatter in the others, the relation for O is the most definite and convincing. Our computed slopes were found to be somewhat smaller but nevertheless compatible with other published results.

\acknowledgments

We thank the TACs at KPNO \& CTIO for granting us observing time,
the local staff there for their assistance, and the IRAF staff for
their ready answers. We appreciate the contributions of Jackie Milingo to the data reductions and abundance analysis for a large fraction of the sample early-on. We are extremely grateful to M.J. Barlow and
X.-W. Liu and to D. Beintema and J.~Bernard-Salas for providing the ISO SWS
line fluxes, to Walter Maciel for sending us his PN chlorine data
in electronic form, to Georges Meynet for responding to our numerous questions regarding nucleosynthesis, to Don Garnett for reading and offering suggestions for improvement on an earlier version of the manuscript, and to an anonymous referee for making numerous suggestions for improving the paper.   We also thank the following students for their
assistance with various calculations related to this project: Lissa Ong
(Williams '04) and Davis Stevenson (Williams '04), both supported by a
grant from the Bronfman Science Center at Williams College; and Megan
Roscioli (Haverford '05) a Keck Summer Fellow at Williams College
supported by the Keck Northeast Astronomy Consortium. KBK also
acknowledges support from the Bronfman Science Center and the Dean of
the Faculty at Williams College. Finally, we thank the NSF for support for this entire project under grant AST-9819123 to Williams College and AST-0307118 to the University of Oklahoma.

\section*{APPENDIX A}

Ten objects whose line strengths were measured and reported as part of an earlier series of papers were added to the current sample because abundances of S, Cl, and Ar were not previously computed. Table~8a gives the resulting ion abundances, ICFs, electron temperatures and densities for these objects. Line strength sources are provided in the table footnotes. The elemental abundances for these objects are provided in Table~8b. Note that electron temperatures deemed very unreliable have been set off with parentheses in Table~8a. 

\section*{APPENDIX B}

The values S$^{+3}$/H$^+$ in column~3 of Table~6 which are given for IC3568, IC4593, NGC~6210, NGC~6572, NGC~6884, and NGC~7027 were calculated from line strength data provided in Table~1 of Dinerstein (1980b). We first used our own values for H$\alpha$ and [N~II] line strengths to derive H$\alpha$ from Dinerstein's H$\alpha$+[N~II] value. We then multiplied the ratio of her [S~IV] 10.5$\mu$m:H$\alpha$ by our observed H$\alpha$/H$\beta$ ratio to give the [S~IV]/H$\beta$ ratio uncorrected for reddening. This ratio was then multiplied by 10$^{cf}$ to correct for reddening, where we took f=-1.09 and $c$ was taken from our own observations. This dereddened [S~IV]/H$\beta$ ratio was then used in a 5-level atom calculation along with our observed [O~III] temperatures and [S~II] densities to finally derive the S$^{+3}$/H$+$ ratio. Collision strengths were taken from Tayal (2000).

\clearpage

\begin{deluxetable}{lcccccccccc}
\tabletypesize{\scriptsize}
\setlength{\tabcolsep}{0.03in}
\tablecolumns{11}
\tablewidth{0in}
\tablenum{1}
\tablecaption{Abundance Compilation}
\tablehead{
\colhead{Object} & 
\colhead{Type} &
\colhead{He/H} &
\colhead{O/H($\times 10^4$)} &
\colhead{Ne/O} &
\colhead{S/O($\times 10^1$)} &
\colhead{Cl/O($\times 10^3$)} &
\colhead{Ar/O($\times 10^2$)} &
\colhead{N/O} &
\colhead{Dist.(kpc)\tablenotemark{a}} &
\colhead{Ref\tablenotemark{b}}
}
\startdata
BB1&halo&0.09 ($\pm$.03)&0.78 ($\pm$.11)&1.63 ($\pm$.23)&0.01 ($\pm$.01)&\nodata&0.02 ($\pm$.01)&0.64 ($\pm$.11)&16.5&3 \\
Cn 2-1&II&0.13 ($\pm$.04)&7.14 ($\pm$1.0)&0.23 ($\pm$.03)&0.09 ($\pm$.01)&0.25 ($\pm$.04)&0.38 ($\pm$.05)&0.39 ($\pm$.06)&2.4& 2\\
DdDm 1&halo&0.10 ($\pm$.03)&1.38 ($\pm$.20)&0.17 ($\pm$.02)&0.18 ($\pm$.06)&0.15 ($\pm$.05)&0.37 ($\pm$.12)&0.21 ($\pm$.03)&11.4&4 \\
Fg 1&II&0.13 ($\pm$.04)&3.73 ($\pm$.53)&0.29 ($\pm$.04)&0.12 ($\pm$.02)&0.33 ($\pm$.10)&0.57 ($\pm$.08)&0.43 ($\pm$.06)&8.0& 2\\
H 4-1&halo&0.12 ($\pm$.04)&1.99 ($\pm$.28)&0.02 ($\pm$.01)&0.01 ($\pm$.001)&\nodata&0.01 ($\pm$.001)&0.29 ($\pm$.04)&15.1& 3\\
Hb 12&II&0.08 ($\pm$.02)&0.37 ($\pm$.05)&0.27 ($\pm$.04)&0.55 ($\pm$.17)&0.18 ($\pm$.11)&1.78 ($\pm$1.1)&0.45 ($\pm$.08)&13.8& 3\\
He 2-21&II&0.12 ($\pm$.04)&2.99 ($\pm$.42)&0.16 ($\pm$.02)&0.06\tablenotemark{c} ($\pm$.01)&0.28 ($\pm$.04)&0.36 ($\pm$.05)&0.15 ($\pm$.02)&11.5&2 \\
He 2-37&II&0.12 ($\pm$.04)&11.0 ($\pm$1.6)&0.23 ($\pm$.03)&0.04 ($\pm$.01)&0.27 ($\pm$.04)&0.43 ($\pm$.06)&0.32 ($\pm$.05)&9.2& 2\\
He 2-48&II&0.11 ($\pm$.03)&4.23 ($\pm$.60)&0.27 ($\pm$.04)&0.05\tablenotemark{c} ($\pm$.02)&0.19 ($\pm$.12)&0.25 ($\pm$.15)&0.31 ($\pm$.04)&8.9& 2\\
He 2-55&II&0.13 ($\pm$.04)&6.87 ($\pm$.97)&0.26 ($\pm$.04)&0.08\tablenotemark{c} ($\pm$.01)&0.53 ($\pm$.32)&0.73 ($\pm$.10)&0.20 ($\pm$.03)&8.2& 2\\
He 2-115&II&0.12 ($\pm$.04)&3.50 ($\pm$.49)&0.13 ($\pm$.02)&0.08 ($\pm$.03)&0.26 ($\pm$.08)&0.63 ($\pm$.09)&0.17 ($\pm$.03)&6.2& 2\\
He 2-123&I&0.15 ($\pm$.05)&6.52 ($\pm$.92)&0.27 ($\pm$.04)&0.16\tablenotemark{c} ($\pm$.02)&0.50 ($\pm$.16)&0.61 ($\pm$.09)&0.81 ($\pm$.11)&6.0& 2\\
He 2-138&II&\nodata&4.60 ($\pm$1.5)&\nodata&0.22\tablenotemark{c} ($\pm$.07)&\nodata&\nodata&0.35 ($\pm$.14)&6.2& 2\\
He 2-140&II&0.08 ($\pm$.02)&3.82 ($\pm$.54)&0.09 ($\pm$.05)&0.34 ($\pm$.05)&0.46 ($\pm$.28)&0.58 ($\pm$.08)&0.38 ($\pm$.06)&5.3& 2\\
He 2-141&II&0.12 ($\pm$.04)&7.64 ($\pm$1.1)&0.19 ($\pm$.03)&0.04\tablenotemark{c} ($\pm$.01)&0.34 ($\pm$.05)&0.33 ($\pm$.04)&0.34 ($\pm$.05)&6.0& 2\\
He 2-157&II&0.12 ($\pm$.04)&1.69 ($\pm$.24)&\nodata&0.38\tablenotemark{c} ($\pm$.05)&0.39 ($\pm$.06)&0.68 ($\pm$.10)&0.40 ($\pm$.06)&3.5& 2\\
He 2-158&II&0.12 ($\pm$.04)&3.24 ($\pm$.46)&0.22 ($\pm$.03)&0.13 ($\pm$.02)&0.25 ($\pm$.15)&0.46 ($\pm$.07)&0.31 ($\pm$.04)&11.1& 2\\
Hu 2-1&II&0.10 ($\pm$.03)&2.43 ($\pm$.34)&0.13 ($\pm$.02)&0.06 ($\pm$.01)&0.17 ($\pm$.05)&0.55 ($\pm$.08)&0.26 ($\pm$.05)&7.2& 3\\
IC418&II&0.07 ($\pm$.02)&1.39 ($\pm$.20)&0.05 ($\pm$.01)&0.53 ($\pm$.17)&\nodata&0.80 ($\pm$.25)&0.59 ($\pm$.08)&9.3& 4\\
IC1297&II&0.13 ($\pm$.04)&7.20 ($\pm$1.0)&0.24 ($\pm$.03)&0.10 ($\pm$.01)&0.27 ($\pm$.04)&0.37 ($\pm$.05)&0.31 ($\pm$.04)&5.2& 2\\
IC2165&II&0.09 ($\pm$.03)&3.11 ($\pm$.44)&0.22 ($\pm$.03)&0.09 ($\pm$.01)&0.22 ($\pm$.03)&0.51 ($\pm$.07)&0.43 ($\pm$.06)&10.5& 3\\
IC2448&II&0.12 ($\pm$.04)&3.25 ($\pm$2.0)&0.21 ($\pm$.13)&0.21\tablenotemark{c} ($\pm$.13)&0.21 ($\pm$.13)&0.38 ($\pm$.23)&0.29 ($\pm$.18)&8.3& 2\\
IC2501&II&\nodata&4.25 ($\pm$.60)&0.26 ($\pm$.04)&0.06 ($\pm$.01)&0.22 ($\pm$.03)&0.41 ($\pm$.06)&0.34 ($\pm$.05)&8.4& 2\\
IC2621&I&0.12 ($\pm$.04)&4.55 ($\pm$.64)&0.23 ($\pm$.03)&0.14 ($\pm$.02)&0.33 ($\pm$.05)&0.87 ($\pm$.12)&0.84 ($\pm$.15)&7.9& 2\\
IC3568&II&0.12 ($\pm$.04)&3.77 ($\pm$.53)&0.19 ($\pm$.03)&0.03 ($\pm$.01)&0.09 ($\pm$.03)&0.28 ($\pm$.09)&0.05 ($\pm$.01)&10.2& 4\\
IC4593&II&0.10 ($\pm$.03)&4.98 ($\pm$.70)&0.18 ($\pm$.03)&0.08 ($\pm$.03)&0.18 ($\pm$.06)&0.37 ($\pm$.05)&0.06 ($\pm$.01)&6.9& 4\\
IC4776&II&0.11 ($\pm$.03)&4.58 ($\pm$.65)&0.22 ($\pm$.03)&0.14 ($\pm$.02)&0.23 ($\pm$.03)&0.36 ($\pm$.22)&0.33 ($\pm$.05)&5.0& 2\\
IC5217&II&0.11 ($\pm$.03)&3.72 ($\pm$.53)&0.23 ($\pm$.03)&0.14 ($\pm$.02)&0.36 ($\pm$.11)&0.45 ($\pm$.06)&0.32 ($\pm$.06)&9.4& 1\\
J320&II&0.11 ($\pm$.03)&2.49 ($\pm$.35)&0.21 ($\pm$.03)&0.08\tablenotemark{c} ($\pm$.01)&0.29 ($\pm$.04)&0.38 ($\pm$.05)&0.52 ($\pm$.07)&13.9& 2\\
J900&II&0.10 ($\pm$.03)&3.54 ($\pm$.50)&0.25 ($\pm$.04)&0.06 ($\pm$.02)&0.11 ($\pm$.02)&0.34 ($\pm$.05)&0.29 ($\pm$.04)&11.8& 3\\
K648&halo&0.10 ($\pm$.03)&0.71 ($\pm$.10)&0.14 ($\pm$.02)&0.03 ($\pm$.02)&\nodata&0.06 ($\pm$.01)&0.04 ($\pm$.01)&10.2& 3\\
M 1-5&II&0.11 ($\pm$.03)&1.29 ($\pm$.18)&0.09 ($\pm$.01)&0.20 ($\pm$.03)&0.38 ($\pm$.12)&0.77 ($\pm$.24)&0.45 ($\pm$.08)&13.5& 2\\
M 1-25&II&0.15 ($\pm$.05)&5.03 ($\pm$.71)&0.07 ($\pm$.01)&0.18 ($\pm$.03)&0.40 ($\pm$.06)&0.66 ($\pm$.09)&0.44 ($\pm$.06)&3.7& 2\\
M 1-34&I&0.15 ($\pm$.05)&6.79 ($\pm$.96)&0.40 ($\pm$.06)&0.19 ($\pm$.03)&0.33 ($\pm$.05)&0.43 ($\pm$.14)&0.66 ($\pm$.14)&2.6& 2\\
M 1-38&II&\nodata&5.91 ($\pm$.55)&\nodata&0.12\tablenotemark{c} ($\pm$.04)&\nodata&0.03 ($\pm$.02)&0.21 ($\pm$.07)&2.0& 2\\
M 1-50&II&0.12 ($\pm$.04)&6.40 ($\pm$.91)&0.21 ($\pm$.03)&0.07 ($\pm$.01)&0.30 ($\pm$.18)&0.38 ($\pm$.05)&0.18 ($\pm$.03)&4.3& 1,2\tablenotemark{d}\\
M 1-54&I&0.15 ($\pm$.05)&5.71 ($\pm$.81)&0.39 ($\pm$.06)&0.21 ($\pm$.03)&0.35 ($\pm$.11)&0.45 ($\pm$.06)&1.06 ($\pm$.15)&4.8& 1,2\tablenotemark{d}\\
M 1-57&I&0.13 ($\pm$.04)&6.42 ($\pm$.91)&0.21 ($\pm$.03)&0.16\tablenotemark{c} ($\pm$.02)&0.42 ($\pm$.13)&0.74 ($\pm$.10)&1.02 ($\pm$.14)&5.3& 1\\
M 1-74&I&0.12 ($\pm$.04)&4.74 ($\pm$.67)&0.25 ($\pm$.04)&0.30 ($\pm$.04)&0.22 ($\pm$.13)&0.69 ($\pm$.10)&0.91 ($\pm$.13)&7.5& 1\\
M 1-80&II&0.10 ($\pm$.03)&8.79 ($\pm$1.2)&0.21 ($\pm$.03)&0.05 ($\pm$.01)&0.06 ($\pm$.04)&0.29 ($\pm$.04)&0.39 ($\pm$.06)&11.3& 1\\
M 2-10&II&0.13 ($\pm$.04)&6.05 ($\pm$.86)&0.25 ($\pm$.04)&0.18\tablenotemark{c} ($\pm$.03)&0.45 ($\pm$.27)&0.57 ($\pm$.08)&0.50 ($\pm$.07)&2.5& 2\\
M 3-4&II&0.15 ($\pm$.05)&5.16 ($\pm$.73)&0.34 ($\pm$.05)&0.03 ($\pm$.01)&0.16 ($\pm$.05)&0.28 ($\pm$.04)&0.37 ($\pm$.05)&17.1& 2\\
M 3-6&II&0.12 ($\pm$.04)&5.59 ($\pm$.79)&0.24 ($\pm$.03)&0.13 ($\pm$.02)&0.21 ($\pm$.13)&0.55 ($\pm$.08)&0.10 ($\pm$.01)&9.9& 2\\
M 3-15&II&0.13 ($\pm$.04)&7.60 ($\pm$1.1)&0.20 ($\pm$.03)&0.11\tablenotemark{c} ($\pm$.02)&0.33 ($\pm$.20)&0.42 ($\pm$.06)&0.36 ($\pm$.05)&7.0& 1,2\tablenotemark{d}\\
NGC 650&II&0.11 ($\pm$.03)&7.11 ($\pm$1.0)&0.39 ($\pm$.06)&0.13 ($\pm$.02)&0.06 ($\pm$.02)&0.52 ($\pm$.07)&0.54 ($\pm$.08)&9.6& 3\\
NGC 1535&II&0.08 ($\pm$.02)&2.66 ($\pm$.38)&0.27 ($\pm$.04)&0.25 ($\pm$.08)&0.16 ($\pm$.02)&0.36 ($\pm$.05)&0.20 ($\pm$.03)&10.1& 3\\
NGC 2022&II&0.09 ($\pm$.03)&7.50 ($\pm$1.1)&0.19 ($\pm$.03)&0.10 ($\pm$.03)&0.61 ($\pm$.19)&1.24 ($\pm$.18)&0.12 ($\pm$.02)&11.0& 3\\
NGC 2371&II&0.10 ($\pm$.03)&7.71 ($\pm$1.1)&0.23 ($\pm$.03)&0.10 ($\pm$.03)&0.61 ($\pm$.09)&1.32 ($\pm$.19)&0.39 ($\pm$.06)&10.5& 3\\
NGC 2392&II&0.08 ($\pm$.02)&3.75 ($\pm$.53)&0.28 ($\pm$.04)&0.15 ($\pm$.09)&\nodata&0.43 ($\pm$.14)&0.30 ($\pm$.04)&9.8& 4\\
NGC 2438&II&0.09 ($\pm$.03)&5.24 ($\pm$.74)&0.32 ($\pm$.05)&0.13 ($\pm$.04)&\nodata&0.42 ($\pm$.06)&0.43 ($\pm$.06)&9.9& 3\\
NGC 2440&I&0.10 ($\pm$.03)&5.19 ($\pm$.73)&0.21 ($\pm$.03)&0.05 ($\pm$.01)&0.53 ($\pm$.17)&0.78 ($\pm$.11)&2.05 ($\pm$.29)&9.5& 3\\
NGC 2792&II&0.11 ($\pm$.03)&7.65 ($\pm$2.4)&0.16 ($\pm$.05)&0.07\tablenotemark{c} ($\pm$.02)&0.68 ($\pm$.22)&0.84 ($\pm$.27)&0.18 ($\pm$.11)&8.8& 2\\
NGC 2867&II&0.12 ($\pm$.04)&5.23 ($\pm$.74)&0.21 ($\pm$.03)&0.06 ($\pm$.01)&0.28 ($\pm$.04)&0.37 ($\pm$.05)&0.26 ($\pm$.04)&8.4& 2\\
NGC 3195&II&0.13 ($\pm$.04)&6.88 ($\pm$.97)&0.40 ($\pm$.06)&0.15 ($\pm$.02)&0.23 ($\pm$.14)&0.40 ($\pm$.06)&0.45 ($\pm$.06)&7.7& 2\\
NGC 3211&II&0.11 ($\pm$.03)&8.38 ($\pm$1.2)&0.16 ($\pm$.02)&0.07\tablenotemark{c} ($\pm$.01)&0.54 ($\pm$.08)&0.76 ($\pm$.11)&0.20 ($\pm$.03)&8.2& 2\\
NGC 3242&II&0.11 ($\pm$.03)&4.10 ($\pm$.58)&0.21 ($\pm$.03)&0.05 ($\pm$.01)&0.26 ($\pm$.04)&0.36 ($\pm$.05)&0.12 ($\pm$.04)&8.7& 2\tablenotemark{e}\\
NGC 3587&II&0.10 ($\pm$.03)&4.53 ($\pm$.64)&0.27 ($\pm$.04)&0.08 ($\pm$.01)&\nodata&0.24 ($\pm$.15)&0.21 ($\pm$.03)&9.3& 1\\
NGC 3918&II&0.11 ($\pm$.03)&5.54 ($\pm$.78)&0.19 ($\pm$.03)&0.05 ($\pm$.01)&0.36 ($\pm$.05)&0.61 ($\pm$.09)&0.39 ($\pm$.06)&8.1& 3\\
NGC 5307&II&0.10 ($\pm$.03)&3.85 ($\pm$.54)&0.23 ($\pm$.03)&0.05 ($\pm$.03)&0.18 ($\pm$.06)&0.35 ($\pm$.05)&0.13 ($\pm$.02)&6.9& 2\\
NGC 5882&II&0.12 ($\pm$.04)&5.48 ($\pm$.77)&0.27 ($\pm$.04)&0.13 ($\pm$.02)&0.31 ($\pm$.04)&0.51 ($\pm$.07)&0.33 ($\pm$.05)&7.1& 3\\
NGC 6210&II&0.11 ($\pm$.03)&5.21 ($\pm$.74)&0.25 ($\pm$.04)&0.13 ($\pm$.04)&0.27 ($\pm$.04)&0.34 ($\pm$.05)&0.22 ($\pm$.03)&7.6& 4\\
NGC 6309&II&0.13 ($\pm$.04)&6.58 ($\pm$.93)&0.22 ($\pm$.03)&0.08 ($\pm$.03)&0.60 ($\pm$.36)&0.82 ($\pm$.12)&0.28 ($\pm$.04)&6.3& 1,2\tablenotemark{d}\\
NGC 6439&I&0.14 ($\pm$.04)&6.32 ($\pm$.89)&0.27 ($\pm$.04)&0.17 ($\pm$.02)&0.44 ($\pm$.14)&0.53 ($\pm$.07)&0.80 ($\pm$.11)&4.0& 1,2\tablenotemark{d}\\
NGC 6563&II&0.12 ($\pm$.04)&6.37 ($\pm$.90)&0.33 ($\pm$.05)&0.06\tablenotemark{c} ($\pm$.01)&0.05 ($\pm$.03)&0.30 ($\pm$.04)&0.32 ($\pm$.05)&6.2& 2\\
NGC 6565&II&0.12 ($\pm$.04)&7.25 ($\pm$1.0)&0.33 ($\pm$.05)&0.13 ($\pm$.02)&0.26 ($\pm$.04)&0.33 ($\pm$.05)&0.41 ($\pm$.06)&4.2& 2\\
NGC 6567&II&0.10 ($\pm$.03)&2.67 ($\pm$.38)&0.18 ($\pm$.03)&0.06 ($\pm$.01)&0.18 ($\pm$.06)&0.21 ($\pm$.03)&0.23 ($\pm$.04)&6.0& 3\\
NGC 6572&I&0.13 ($\pm$.04)&4.29 ($\pm$.61)&0.23 ($\pm$.03)&0.10 ($\pm$.01)&0.23 ($\pm$.07)&0.53 ($\pm$.07)&0.67 ($\pm$.12)&7.7& 1\\
NCG 6578&II&0.12 ($\pm$.04)&7.42 ($\pm$1.0)&0.30 ($\pm$.04)&0.11 ($\pm$.02)&0.25 ($\pm$.08)&0.43 ($\pm$.06)&0.31 ($\pm$.04)&6.1& 3\\
NGC 6629&II&0.11 ($\pm$.03)&4.50 ($\pm$.64)&0.20 ($\pm$.03)&0.07 ($\pm$.01)&0.21 ($\pm$.03)&0.42 ($\pm$.06)&0.14 ($\pm$.02)&6.7& 2\\
NGC 6720&II&0.12 ($\pm$.04)&7.64 ($\pm$1.1)&0.27 ($\pm$.04)&0.06 ($\pm$.02)&0.32 ($\pm$.05)&0.56 ($\pm$.08)&0.35 ($\pm$.05)&8.1&4\\
NGC 6790&II&0.12 ($\pm$.04)&3.36 ($\pm$.48)&0.17 ($\pm$.02)&0.06 ($\pm$.01)&0.18 ($\pm$.11)&0.26 ($\pm$.04)&0.18 ($\pm$.03)&6.3& 1\\
NGC 6826&II&0.11 ($\pm$.03)&4.16 ($\pm$.59)&0.20 ($\pm$.03)&0.07 ($\pm$.02)&0.20 ($\pm$.03)&0.36 ($\pm$.11)&0.12 ($\pm$.02)&8.5& 4\\
NGC 6879&II&0.11 ($\pm$.03)&3.79 ($\pm$.54)&0.22 ($\pm$.03)&0.09 ($\pm$.03)&0.15 ($\pm$.09)&0.39 ($\pm$.06)&0.18 ($\pm$.03)&7.5& 1\\
NGC 6884&II&0.12 ($\pm$.04)&5.55 ($\pm$.78)&0.22 ($\pm$.03)&0.11 ($\pm$.02)&0.35 ($\pm$.11)&0.54 ($\pm$.08)&0.41 ($\pm$.07)&5.9& 1\\
NGC 6886&II&0.12 ($\pm$.04)&5.24 ($\pm$.74)&0.23 ($\pm$.03)&0.11 ($\pm$.02)&0.42 ($\pm$.13)&0.78 ($\pm$.11)&0.43 ($\pm$.08)&7.5& 1\\
NGC 6891&II&0.11 ($\pm$.03)&4.25 ($\pm$.60)&0.19 ($\pm$.03)&0.04 ($\pm$.01)&0.16 ($\pm$.05)&0.39 ($\pm$.24)&0.10 ($\pm$.02)&7.3& 1\\
NGC 7009&II&0.12 ($\pm$.04)&5.63 ($\pm$.80)&0.23 ($\pm$.03)&0.15 ($\pm$.05)&0.32 ($\pm$.05)&0.47 ($\pm$.07)&0.42 ($\pm$.06)&7.8& 4\\
NGC 7026&I&0.14 ($\pm$.04)&7.31 ($\pm$1.0)&0.30 ($\pm$.04)&0.21 ($\pm$.03)&0.50 ($\pm$.16)&0.73 ($\pm$.10)&0.76 ($\pm$.11)&8.7& 1\\
NGC 7027&II&0.10 ($\pm$.03)&4.15 ($\pm$.59)&0.17 ($\pm$.02)&0.13 ($\pm$.02)&0.41 ($\pm$.06)&0.65 ($\pm$.09)&0.56 ($\pm$.08)&8.5& 3\\
NGC 7293&II&0.13 ($\pm$.04)&6.47 ($\pm$.91)&0.65 ($\pm$.09)&0.09 ($\pm$.03)&\nodata&0.48 ($\pm$.07)&0.36 ($\pm$.05)&8.2& 4\\
NGC 7662&II&0.10 ($\pm$.03)&4.19 ($\pm$.59)&0.17 ($\pm$.02)&0.10 ($\pm$.01)&0.50 ($\pm$.07)&0.59 ($\pm$.08)&0.18 ($\pm$.03)&8.9& 3\\
PB 6&I&0.17 ($\pm$.05)&6.77 ($\pm$.96)&0.24 ($\pm$.03)&0.03 ($\pm$.004)&0.27 ($\pm$.04)&0.87 ($\pm$.12)&1.13 ($\pm$.16)&9.0& 2\\
PC 14&II&0.12 ($\pm$.04)&7.88 ($\pm$1.1)&0.27 ($\pm$.04)&0.09 ($\pm$.01)&0.26 ($\pm$.16)&0.38 ($\pm$.05)&0.18 ($\pm$.03)&4.4& 2\\
Pe 1-18&I&0.14 ($\pm$.04)&4.56 ($\pm$.64)&0.21 ($\pm$.03)&0.21 ($\pm$.03)&0.40 ($\pm$.13)&0.72 ($\pm$.10)&1.70 ($\pm$.30)&3.9& 1,2\tablenotemark{d}\\
Th 2-A&II&0.13 ($\pm$.04)&6.23 ($\pm$.88)&0.31 ($\pm$.04)&0.07 ($\pm$.02)&0.12 ($\pm$.02)&0.52 ($\pm$.32)&0.49 ($\pm$.07)&7.2& 2 \\
\cutinhead{Averages\tablenotemark{f}}
Type I & \nodata & 0.137$\pm$.018 (12) & 5.76$\pm$1.06 (12) & 0.27$\pm$.066 (12) & 0.16$\pm$.075 (12) & 0.38$\pm$.11 (12) & 0.66$\pm$.15 (12) & 1.03$\pm$.49 (12) & \nodata & \nodata \\
Type II &\nodata& 0.113$\pm$.016 (66) & 5.09$\pm$2.02 (69) & 0.23$\pm$.085 (66) & 0.12$\pm$.097 (69) & 0.29$\pm$.14 (62) & 0.50$\pm$.27 (68) & 0.31$\pm$.14 (69) & \nodata & \nodata \\
Halo & \nodata & 0.103$\pm$.011 (4) & 1.00$\pm$.64 (5) & 0.41$\pm$.61 (5) & 0.058$\pm$.071 (4) & 0.13$\pm$.11 (2) & 0.12$\pm$.15 (4) & 0.33$\pm$.27 (4) & \nodata & \nodata \\
All & \nodata & 0.115$\pm$.018 (82) & 5.01$\pm$2.04 (86) & 0.25$\pm$.17 (83) & 0.12$\pm$.095 (85) & 0.30$\pm$.14 (76) & 0.51$\pm$.27 (84) & 0.41$\pm$.35 (85) & \nodata & \nodata 
\enddata

\tablenotetext{a}{Galactocentric distances were calculated from heliocentric distances given by Maciel (1984) for H4-1, He2-48, He2-55, and M3-15; Cahn et al (1992) for DdDm-1 and Hu2-1; Sabbadin (1986) for K648; Torres-Peimbert et al. (1990) for BB1; and Zhang (1995) for the rest. A solar galactocentric distance of 8.5~kpc was assumed.}

\tablenotetext{b}{Abundance references: 1=Kwitter \& Henry (2001); 2=Milingo, Henry, \& Kwitter (2002); 3=Kwitter, Henry, \& Milingo (2003); 4=Appendix~A, this paper.}

\tablenotetext{c}{S/O value differs slightly from the one(s) originally published in the earlier reference(s) due to an improvement in the estimated electron temperature relevant for this object. This value supersedes the former one.}
\tablenotetext{d}{Values quoted are averages of abundances from the two sources}

\tablenotetext{e}{Values quoted are averages of positions A and B abundances}

\tablenotetext{f}{The values in parentheses indicate the number of sample objects included in the average.}

\end{deluxetable}

\clearpage

\begin{deluxetable}{lcccccc}
\tabletypesize{\scriptsize}
\setlength{\tabcolsep}{0.03in}
\tablecolumns{7}
\tablewidth{0in}
\tablenum{2}
\tablecaption{Electron Temperatures and Densities\tablenotemark{a}}
\tablehead{
\colhead{Object} & 
\colhead{T$_{[O~III]}$ (K)} &
\colhead{T$_{[N~II]}$ (K)} &
\colhead{T$_{[O~II]}$ (K)} &
\colhead{T$_{[S~II]}$ (K)} &
\colhead{T$_{[S~III]}$ (K)} &
\colhead{N$_e$ (cm$^{-3}$)} 
}
\startdata
BB1 &  12400 &  10500 &   8000 &      \nodata &      \nodata &  7100\tablenotemark{c}\\
Cn 2-1 &   9600 &  10200 &  13700 &  13300 &  12000 &  4100\\
DdDm1 &  11700 &  11400 & 10100 & 7900 &  12700 & 4000 \\
Fg 1&   9300 &   8300 &   7100 &      \nodata &   9400 &  600\\
H4-1 &  12300 &  10200 &   6100 &      \nodata &      \nodata &  400\\
Hb 12 &  18500 &   9300 &      \nodata &      \nodata &      \nodata &  6300\\
He 2-21 &  12200 &   7600 &  11500 &      \nodata &  17600 &  1500\\
He 2-37 &  12000 &  10000 &  14500 &  10300 &  13800 &  200\\
He 2-48 &  11200 &   9600 &      \nodata &   6500 &      \nodata &  10\\
He 2-55 &  12200 &   9500 &  16100 &      \nodata &  17200 &  200\\
He 2-115 &   9100 &  10100 &  12400 &  16300 &  11500 &  11700\\
He 2-123 &   6200 &   6600 &   5800 &   6400 &   7700 &  2000\\
He 2-138 &   5800 &   5600 &   5900 &   6300 &      \nodata &  5300\\
He 2-140 &   7000 &   7100 &   7100 &   6100 &   8200 &  5300\\
He 2-141 &  12400 &   8800 &   7300 &      \nodata &  15500 &  1400\\
He 2-157 &  10500 &   9900 &   7400 &      \nodata &   9800 &  5200\\
He 2-158 &   9200 &  10000 &   7800 &  15200 &   9700 &  2400\\
Hu 2-1 &   9100 &  11700 &  20600 &      \nodata &  11100 &  8500\\
IC418 &  8500 &  10700 & 17100 & 8700 &  8100 & 3300 \\
IC1297 &   9900 &   8900 &   8100 &      \nodata &  11000 &  2400\\
IC2165 &  13000 &  12700 &   9800 &  10300 &  14200 &  4100\\
IC2448 &  12500 &      \nodata &      \nodata &      \nodata &      \nodata &  10\\
IC2501 &   9500 &  11200 &  10600 &  11700 &  12700 &  4800\\
IC2621 &  12500 &  11200 &      \nodata &      \nodata &  12800 &  8300\\
IC3568 &  10400 &  6700 & 8800 & \nodata &  10300 & 800 \\
IC4593 &  8100 &  7300 & 6800 & \nodata &  9000 & 1700 \\
IC4776 &   9500 &  13500 &  15600 &  15000 &  11500 &  4000\\
IC5217&  11000 &  10700 &  12300 &   8800 &  11200 & 10000  \\
J320 &  12100 &   9900 &  16100 &   6100 &  17600 &  4600\\
J900 &  11600 &  11500 &  10300 &   7800\tablenotemark{b} &  13000 &  3600\\
K648 &  11800 &   9200 &   8400 &      \nodata &      \nodata &  1000\\
M1-5 &  10000 &  10100 &      \nodata &  11700 &  10100 &  6400\\
M1-25 &   8400 &   8200 &   7900 &   9100 &   9300 &  4300\\
M1-34 &   9200 &   8200 &   5400 &   9400 &   9100 &  700\\
M1-38 &   6300 &   6300 &   5900 &   6600 &   8500 &  5100\\
M1-50 &  10300 &  10400 &  11700 &      \nodata &  12900 &  4500\\
M1-54 &   9800 &   8700 &   6200 &  11200 &   9200 &  1500\\
M1-57 &  12300 &  11200 &  13000 &   6700 &  15800 & 4500  \\
M1-74 &   9300 &  14700 &      \nodata &      \nodata &  11400 &  2100 \\
M1-80 &   9500 &  10000 &   6200 &  10400 &  13500 &  800\\
M2-10 &   6300 &   6700 &   5100 &  11400 &   7200 &  1300\\
M3-4 &  11500 &   9300 &  14000 &   8500 &  11000 &  200\\
M3-6 &   8000 &   8700 &   9500 &      \nodata &   8400 &  1700\\
M3-15 &   8000 &    10400 &  10800 &   9000 &  10500 &  3700\\
NGC 650 &  10800 &   9700 &  14800 &  10300 &  13300\tablenotemark{b} &  200\\
NGC 1535 &  11400 &      \nodata &   7400 &      \nodata &  17900\tablenotemark{c} &  300\\
NGC 2022 &  13600 &  10300 &   7500 &      \nodata &  19600 &  800\\
NGC 2371 &  12100 &   8600\tablenotemark{b} &   6800 &  11400\tablenotemark{c} &  13300 &  1000\\
NGC 2392 &  11900 &  11400 & 6800 & \nodata &  10400 & 2000 \\
NGC 2438 &  10300 &  10400\tablenotemark{b} &  13100 &   9700\tablenotemark{c} &  10000\tablenotemark{c} &  200\\
NGC 2440 &  12600 &  10300 &   8300 &  13800\tablenotemark{c} &  13800\tablenotemark{b} &  1800\\
NGC 2792 &  13700 &  10200 &   7900 &      \nodata &      \nodata &  2800\\
NGC 2867 &  11200 &  10100 &   8200 &      \nodata &  12800 &  2100\\
NGC 3195 &   8900 &   8000 &  11300 &  12500 &   9000 &  200\\
NGC 3211 &  13300 &  10300 &   6700 &      \nodata &      \nodata &  1200\\
NGC 3242 &  11200 &  12100 &   9600 &      \nodata &  15800 &  2100\\
NGC~3587 &  10600 &   9400 &  11600 &      \nodata &  12400 &  100\\
NGC 3918 &  12200 &  10400 &   8500 &   8600 &  15300 &  3800\\
NGC 5307 &  12300 &  11000 &   9900 &      \nodata &      \nodata &  2600\\
NGC 5882 &   9100 &   9900 &   9900 &      \nodata &  10100 &  2200\tablenotemark{b}\\
NGC 6210 &  9400 &  10200 & 10100 & 9200 &  9500 & 4100 \\
NGC-6309 &  11600 &   8900 &  10100 &  12000 &  11400 &  3600\\
NGC~6439 &   9700 &   9500 &   9000 &   6700 &  10000 &  4100\\
NGC 6563 &  10300 &   9000 &  13100 &      \nodata &      \nodata &  100\\
NGC 6565 &  10100 &   9400 &   6100 &  11800 &  10300 &  1300\\
NGC 6567 &  11000 &  12600\tablenotemark{b} &  11800 &  16300\tablenotemark{c} &  14500 &  6700\tablenotemark{c}\\
NGC~6572 &  10100 &  12300 &      \nodata &      \nodata &  11500 & 9500 \\
NGC 6578 &   7800 &  10000 &   9000 &      \nodata &   9300 &  2400\tablenotemark{b}\\
NGC 6629 &   8500 &  10300 &   6500 &      \nodata &   9200 &  1100\\
NGC 6720 &  11050 &  9400 & 5800 & 12900 &  11400 & 600 \\
NGC~6790 &  12400 &  18300 &      \nodata &      \nodata &  14800 & $>$10000 \\
NGC 6826 &  8900 &  8700 & 6700 & \nodata &  9900 & 1700 \\
NGC~6879 &  10100 &   8900 &  12700 &  15800 &  10400 & 7400\\
NGC~6884 &  10400 &   9900 &  11600 &  14400 &  10500 & 6500 \\
NGC~6886 &  12100 &   9800 &  10300 &   8200 &  12900 &  7200\\
NGC~6891 &   9200 &  10600 &   9200 &      \nodata &  11500 &  10000\\
NGC 7009 &  9500 &  10200 & 10600 & 13900 &  10600 & 4300 \\
NGC~7026 &   8500 &   9400 &   7800 &   6400 &   9500 &  3300\\
NGC 7027 &  13600 &  20100 &      \nodata &      \nodata &  15100 & 81000 \\
NGC 7293 &  9100 &  8500 & 8600 & 13600 &  5100 & 100 \\
NGC 7662 &  12700 &  10300 &  12200 &      \nodata &  13600 &  2700\\
PB 6 &  14600 &   9800 &   7600 &      \nodata &  16300 &  2200\\
PC 14 &   8800 &   8200 &   7400 &   6900 &  10200 &  2200\\
Pe 1-18 &   9800 &  12600 &      \nodata &      \nodata &  12500 &  12100\\
Th 2-A &  11600 &  11700 &   5900 &      \nodata &      \nodata &  1200
\enddata

\tablenotetext{a}{Temperature and density uncertainties are each $\pm$10\%, unless otherwise noted.}
\tablenotetext{b}{Uncertainty estimated to be $\pm$20\%}
\tablenotetext{c}{Uncertainty estimated to be $\pm$30\%}
\end{deluxetable}

\clearpage

\clearpage

\begin{deluxetable}{lcccc}
\tablecolumns{5}
\tablewidth{0in}
\tablenum{3}
\tablecaption{Comparison of Abundance Ratio Averages}
\tablehead{
\colhead{Paper\tablenotemark{a}} & 
\colhead{log(S/O)} &
\colhead{log(Cl/O)} &
\colhead{log(Ar/O)} &
\colhead{Comments}
}
\startdata
\cutinhead{Planetary Nebulae\tablenotemark{b}}
GM88 & -1.87 (11) & \nodata & -1.94 (3) & Galactic\\
FP91 & -1.84 (8) & \nodata & -2.25 (8) & Type I galactic\\
FP92 & -1.95 (12) & \nodata & -1.95 (17) & Type II galactic\\
FP93a & -1.63 (15) & \nodata & -2.47 (15) & non-Type I in LMC\\
FP93b & -1.57 (8) & \nodata & -2.16 (8) & Type I in LMC\\
KB94 & -1.77 (43) & \nodata & -2.32 (42) & Galactic\\
T03 & -1.70 ( 15) & -3.44 ( 14) & -2.29 ( 14) & Galactic\\
This Paper & -1.91 ( 85) & -3.52 (75) & -2.29 (84) & Galactic\\
\cutinhead{H II Regions\tablenotemark{b}}
R99 & -1.45 & -3.30 & -2.05 & Ave., 7 Galactic \\
E98 & -1.46 & -3.30 & -1.83 & Orion Nebula \\
P92 & -1.49 & -3.33 & -2.18 & M17 \\
KBG & -1.55 (19) & \nodata & -2.21 (16) & 20 H II Regions in M101 (Ave)\\
L02 & -1.46 & -3.54 & -2.35 & N5461 in M101\\
L02 & -1.57 & -3.67 & -2.38 & N5471 in M101\\
\cutinhead{Sun}
GS98 & -1.49 & -3.41 & -2.29 &
\enddata
\tablenotetext{a}{GM88=Guti{\'e}rrez-Moreno \& Moreno (1988); FP91=Freitas Pacheco, Maciel, Costa, \& Barbuy (1991); FP92=Freitas Pacheco, Maciel, \& Costa (1992); FP93a=Freitas Pacheco, Costa, \& Maciel (1993a); FP93b=Freitas Pacheco, Barbuy, Costa, \& Idiart (1993b);  KB94=Kingsburgh \& Barlow (1994); T03=Tsamis et al. (2003); R99=Rodr{\'i}guez (1999); E98=Esteban et al. (1998; gas phase abundance values from their Table~17); P92=Peimbert et al. (1992); KBG=Kennicutt, Bresolin, \& Garnett (2003); L02=Luridiana et al. (2002); GS98=Grevesse \& Sauval (1998, where meteoritic values are used for S and Cl, photospheric value for Ar, but using the oxygen abundance from Allende Prieto et al. 2001).}
\tablenotetext{b}{Values in parentheses indicate number of survey objects included in the computed average.}
\end{deluxetable}
\clearpage

\begin{deluxetable}{lcccc}
\tablecolumns{5}
\tablewidth{0in}
\tablenum{4}
\tablecaption{12+log(X/H) vs. 12+log(O/H): Fit Parameters}
\tablehead{
\colhead{Quantity} & 
\colhead{Intercept} &
\colhead{Slope} &
\colhead{Corr. Coef.} &
\colhead{Number} 
}
\startdata
12+log(Ne/H) & -2.18$\pm$.83 & 1.18$\pm$.095 & 0.81 & 82 \\
12+log(S/H) & -1.39$\pm$1.17 & 0.93$\pm$.14 & 0.60 & 85 \\
12+log(Cl/H) & -5.27$\pm$1.04 & 1.20$\pm$.12 & 0.76 & 75 \\
12+log(Ar/H) & -5.33$\pm$1.20 & 1.34$\pm$.14 & 0.73 & 84 
\enddata
\end{deluxetable}

\clearpage

\begin{deluxetable}{lcccccc}
\tabletypesize{\scriptsize}
\tablecolumns{7}
\tablewidth{0in}
\tablenum{5}
\tablecaption{Photoionization Models}
\tablehead{
\colhead{} & \multicolumn{2}{c}{IC 4593} & \multicolumn{2}{c}{Hu2-1} & \multicolumn{2}{c}{NGC 3242} \\
\colhead{Line ID/Parameter} & 
\colhead{Obs.\tablenotemark{a}} & 
\colhead{Model} & 
\colhead{Obs.\tablenotemark{b}} & 
\colhead{Model} & 
\colhead{Obs.\tablenotemark{c}} & 
\colhead{Model} 
} 
\startdata
C IV $\lambda$1549 & 5.8 & 1.4 & \nodata & 1.8 & 36 & 115 \\
C III] $\lambda$1909 & 2.5 & 2.8 & 44 & 41 & 235 & 244 \\
\[[O II] $\lambda$3727 & 42 & 49 & 77 & 78 & 9.5 & 13\\
\[[Ne III] $\lambda$3869 & 29 & 28 & 18 & 16 & 102 & 110\\
\[[S II] $\lambda$4072 & 0.49 & 0.28 & 1.4 & 0.59 & 1.1: & 0.11\\
C II $\lambda$4267 & 0.17: & 0.04 & 0.4: & 0.14 &0.75 & 0.25\\
\[[O III] $\lambda$4363 & 1.9 & 1.9 & 2.3 & 2.6 & 13 & 14\\
He I $\lambda$4471 & 4.9 & 5.4 & 4.9 & 4.9 & 3.9  & 4.2\\
He II $\lambda$4686 & 0.67  & 0.32 & 0.2:: & 0.10 & 24 & 27\\
H$\beta$ $\lambda$4861 & 100 & 100 & 100 & 100 & 100 & 100\\
\[[O III] $\lambda$5007 & 556 & 587 & 438 & 434 & 1275 & 1432\\
\[[Cl III] $\lambda$5517 & 0.39 & 0.38 & 0.1:: & 0.16 & 0.30 & 0.44\\
\[[Cl III] $\lambda$5537 & 0.35 & 0.37 & 0.3: & 0.26 & 0.30 & 0.41\\
\[[N II] $\lambda$5755 & 0.09:: & 0.11 & 2.3 & 1.2 & 0.10 & 0.04\\
He I $\lambda$5876 & 15 & 16 & 15 & 15 & 12 & 13\\
\[[O I] $\lambda$6300 & \nodata & 0.01 & 2.3 & 0.03 & \nodata & $<$0.01\\
\[[S III] $\lambda$6312 & 0.73 & 0.58 & 0.6 & 0.52 & 0.85 & 0.54\\
H$\alpha$ $\lambda$6563 & 287 & 288 & 286 & 285 & 278 & 283 \\
\[[N II] $\lambda$6584 & 11 & 12 & 63 & 64 & 2.7: & 2.3\\
\[[S II] $\lambda$6716 & 0.55 & 0.90 & 0.4 & 0.46 & 0.35 & 0.27\\
\[[S II] $\lambda$6731 & 0.78 & 1.3 & 0.9 & 0.92 & 0.45 & 0.37\\
\[[Ar III] $\lambda$7135 & 9.7 & 11 & 10 & 14 & 8.7 & 9.3\\
\[[O II] $\lambda$7325 & 2.4 & 1.7 & 29 & 11 & 1.2 & 0.73\\
\[[Ar III] $\lambda$7751 & 2.3 & 2.6 & 2.4 & 3.3 & 2.0 & 2.2\\
\[[Cl IV] $\lambda$8045 & \nodata & 0.50 & \nodata & 0.30 & 0.90 & 0.50\\
\[[S III] $\lambda$9069 & 16: & 16 & 5.2 & 10 & 6.0 & 7.8\\
\[[S III] $\lambda$9532 & 45: & 41 & 22 & 24 & 18 & 19\\ 
\cutinhead{Model Parameters\tablenotemark{d}}
He/H & 0.10 & 0.11 & 0.11 & 0.10 & 0.11 & 0.11\\
C/H & 3.98E-5 & 3.58E-5 & 4.35E-4 & 1.51E-4 & 3.28E-4 & 4.42E-4 \\
N/H & 2.99E-5 & 5.42E-5 & 6.32E-5 & 8.71E-5 & 4.92E-5 & 4.90E-5\\
O/H & 4.98E-4 & 4.51E-4 & 2.43E-4 & 2.45E-4 & 4.10E-4 & 4.07E-4\\
Ne/H & 8.96E-5 & 6.67E-5 & 3.16E-5 & 3.16E-5 & 8.61E-5 & 6.92E-5\\
S/H & 3.98E-6 & 3.58E-6 & 1.46E-6 & 1.45E-6 & 2.05E-6 & 2.29E-6\\
Cl/H & 8.96E-8 & 1.08E-7 & 4.13E-8 & 4.17E-8 & 1.07E-7 & 1.07E-7\\
Ar/H & 1.84E-6 & 1.64E-6 & 1.34E-6 & 1.35E-6 & 1.48E-6 & 1.51E-6\\
T$_{eff}$(K) & 40,000\tablenotemark{e} & 48,000 & 38,000\tablenotemark{f} & 41,000 & 75,000\tablenotemark{e} & 90,000\\
log L$_{H\beta}$(erg/s) & 34.5\tablenotemark{g} & 34.8 & 34.1\tablenotemark{g} & 34.2 & 34.4 & 34.8\\
N$_e$ (cm$^{-3}$) & 1700 & 1900 & 8500 & 8700 & 1700 & 1800\\
radius (pc) & 0.098\tablenotemark{g} & 0.13 & 0.018\tablenotemark{g} & 0.036 & 0.098\tablenotemark{g} & 0.12\\
filling factor & \nodata & 0.5 & \nodata & 1 & \nodata & 1\\
\enddata
\tablenotetext{a}{Optical and UV line strengths are taken from Kwitter \& Henry (1998)}
\tablenotetext{b}{Optical line strengths are taken from Kwitter, Henry, \& Milingo (2003), while UV line strengths are taken from Henry, Kwitter, \& Howard (1996)}
\tablenotetext{c}{Optical line strengths are taken from Milingo, Kwitter, \& Henry (2002) and are averages of positions A and B; UV line strengths are taken from Henry, Kwitter, \& Bates (2000)}
\tablenotetext{d}{All observed abundances except for carbon are taken from Table~1 of this paper; observed carbon abundances are taken from Henry, Kwitter, \& Bates (2000); observed electron densities are taken from same sources as line strengths.}
\tablenotetext{e}{M{\'e}ndez, Kudritzki, \& Herrero (1992)}
\tablenotetext{f}{Stasi{\'n}ska, G{\'o}rny, \& Tylenda (1997)}
\tablenotetext{g}{Cahn, Kaler, \& Stanghellini (1992)}
\end{deluxetable}

\begin{deluxetable}{lcccc}
\tablecolumns{5}
\tablewidth{0in}
\tablenum{6}
\tablecaption{S$^{+3}$ Abundance Comparisons}
\tablehead{
\colhead{Object}&\colhead{S$^{+3}$/H$^+$(ICF)}&\colhead{S$^{+3}$/H$^+$(IR)}&\colhead{ICF/IR}&\colhead{Reference\tablenotemark{a}}
}
\startdata
DdDm1 & 3.16E-7 & 2.3E-7 & 1.4 & 1,7 \\
H4-1 & 2.18E-8 & $\le$3.8E-7 & $\ge$0.057 & 3,7 \\
Hb 12 &4.03E-7 &3.69E-8 & 11 & 2,2  \\
IC3568 & 4.12E-7 & 1.18E-6 & 0.35 & 1,6 \\
IC4593 & 7.08E-7 & 9.88E-7 & 0.72 & 1,6 \\
NGC 3918 &8.34E-7 &2.70E-6 & 0.31 & 2,2 \\
NGC 5882 &4.00E-6 &4.44E-6 & 0.90 & 2,2 \\
NGC 6210 & 2.21E-6 & 3.48E-6 & 0.64 & 1,6\\
NGC 6567 &7.03E-7 &2.34E-6 & 0.30 & 2,2 \\
NGC 6572 & 2.11E-6 & 6.36E-7 & 3.3 & 5,6 \\
NGC 6578 &4.28E-6 &5.98E-6 & 0.72 & 2,2 \\
NGC 6884 & 2.87E-6 & 3.07E-6 & 0.93 & 5,6 \\
NGC 7027 & 1.96E-6 & 2.27E-6 & 0.86 & 4,6 \\
NGC 7662 &3.20E-6 &2.74E-6 & 1.2 & 2,2 \\
\enddata

\tablenotetext{a}{The first number refers to the source for value in column~2, the second number likewise for column~3. 1=Table~8a, this paper; 2=Table~4a, Paper~III, where these values have been scaled by 0.75 (6.42/8.54) to reflect recent changes in [S~IV] collision strengths published by Tayal (2000); 3=Table~4b, Paper~III; 4=Table~4c, Paper~III; 5=Table~5c, Paper~I; 6=Dinerstein (1980b; these values were updated according to the procedure described in Appendix~B); 7=Dinerstein et al. (2003).} 

\end{deluxetable}

\clearpage

\begin{deluxetable}{lcccccc}
\tablecolumns{7}
\tablewidth{0in}
\tablenum{7}
\tablecaption{Abundance Gradients: Fit Parameters}
\tablehead{
\colhead{Quantity} & 
\colhead{Intercept} &
\colhead{Slope (dex kpc$^{-1}$)} &
\colhead{Corr. Coef.} &
\colhead{Number} &
\colhead{Comp. Slope\tablenotemark{a}} &
\colhead{[X/H]$_{8.5}$\tablenotemark{b}}
}
\startdata
12+log(O/H) & 8.97$\pm$.069 & -0.037$\pm$.008 & -0.45 & 79 & -0.058$\pm$.007 & -0.08\\
12+log(Ne/H) & 8.37$\pm$.11 & -0.044$\pm$.014 & -0.35 & 77 & -0.036$\pm$.010 & -0.06\\
12+log(S/H) & 7.05$\pm$.080 & -0.048$\pm$.0098 & -0.48 &81 & -0.077$\pm$.011 & -0.56\\
12+log(Cl/H) & 5.45$\pm$.11 & -0.045$\pm$.013 & -0.36 &74 & -0.070$\pm$.010 & -0.21\\
12+log(Ar/H) & 6.58$\pm$.079 & -0.030$\pm$.010 & -0.34 & 77 & -0.051$\pm$.010 & -0.19
\enddata
\tablenotetext{a}{Values quoted directly from Maciel \& Quireza (1999) for O, Ne, S, and Ar, and from Maciel \& Chiappini (1994) for Cl.}
\tablenotetext{b}{log(X/H)$_{8.5}$-log(X/H)$_{\odot}$, the logarithmic offset from the solar value at 8.5~kpc galactocentric distance, where the sources for solar values are discussed in the text.}
\end{deluxetable}

\clearpage

\begin{deluxetable}{lrrrrrrrrrr}
\tabletypesize{\scriptsize}
\setlength{\tabcolsep}{0.03in}
\tablecolumns{11}
\tablewidth{0in}
\tablenum{8A}
\tablecaption{Ionic Abundances, Temperatures, \& Densities}
\tablehead{
\colhead{Parameter} & 
 \colhead{DdDm1\tablenotemark{a}}   &
\colhead{IC418\tablenotemark{b}}   &
\colhead{IC3568\tablenotemark{a}} &
\colhead{IC4593\tablenotemark{a}} &
\colhead{NGC6720\tablenotemark{a}} &
\colhead{NGC6826$^{a,e}$}   &
\colhead{NGC7009$^{a,e}$}   &
\colhead{NGC7293$^{c,e}$}   &
\colhead{NGC2392$^{b,d}$}   &
\colhead{NGC6210\tablenotemark{a}}
}
\startdata

He$^+$/H$^+$& 0.10    &  6.93E-02&  0.12    &  0.10    &  7.52E-02&  0.11    &  0.10    &  0.13    &  4.20E-02&  0.11  \\
He$^{+2}$/H$^+$& \nodata    &    \nodata    &  1.01E-03&  5.76E-04&  4.03E-02&    \nodata    &  1.08E-02&  2.94E-03&  3.33E-02&  1.08E-03    \\
ICF(He)&  1.00&      1.00&      1.00&      1.00&      1.00&      1.00&      1.00&      1.00&      1.00&      1.00     \\
O$^o$/H$^+$& 2.79E-06&  7.45E-06&    \nodata    &\nodata    &  1.91E-05& \nodata    &  4.63E-07&  9.83E-05&  3.55E-07&  6.83E-06  \\
O$^+$/H$^+$& 3.78E-05&  6.21E-05&  1.74E-05&  8.81E-05&  9.97E-05&  2.67E-05&  7.54E-06&  3.79E-04&  3.50E-05&  2.80E-05  \\
O$^{+2}$/H$^+$& 1.00E-04&  7.66E-05&  3.57E-04&  4.08E-04&  3.98E-04&  3.90E-04&  5.03E-04&  2.52E-04&  1.74E-04&  4.88E-04  \\
ICF(O)&  1.00&      1.00&      1.01&      1.01&      1.54&      1.00&      1.10&      1.02&      1.79&      1.01 \\
N$^{+}$/H$^+$& 7.84E-06&  3.69E-05&  8.18E-07&  4.98E-06&  3.49E-05&  3.18E-06&  3.16E-06&  1.36E-04&  1.06E-05&  6.17E-06  \\
ICF(N)&  3.65&      2.23&     21.63&      5.66&      7.66&     15.62&     74.65&      1.70&     10.71&     18.58 \\
Ne$^{+2}$/H$^+$& 1.71E-05&  4.16E-06&  6.77E-05&  7.40E-05&  1.08E-04&  7.90E-05&  1.17E-04&  1.65E-04&  4.97E-05&  1.23E-04 \\
ICF(Ne)& 1.38&      1.81&      1.06&      1.22&      1.92&      1.07&      1.12&      2.56&      2.15&      1.07  \\
S$^+$/H$^+$& 2.37E-07&  3.05E-07&  1.35E-08&  9.11E-08&  6.73E-07&  4.87E-08&  1.37E-07&  5.33E-07&  2.67E-07&  2.92E-07 \\
S$^{+2}$/H$^+_{NIR}$&  1.87E-06&  6.53E-06&  7.36E-07&  3.28E-06&  3.09E-06&  1.94E-06&  3.27E-06&  4.76E-06&  3.92E-06&  4.21E-06  \\
S$^{+2}$/H$^+_{6312}$& 2.46E-06&  2.30E-06&  4.60E-06&  7.56E-06&  5.69E-06&  3.09E-06&  3.62E-06&  2.88E-07&  2.87E-06&  3.18E-06 \\
ICF(S)&    1.15&      1.08&      1.55&      1.21&      1.26&      1.43&      2.46&      1.05&      1.33&      1.49   \\
Cl$^{+2}$/H$^+$& 2.13E-08&    \nodata    &  3.25E-08&  9.09E-08&  1.18E-07&  8.51E-08&  9.01E-08&    0.    &    0.    &  8.44E-08  \\
Cl$^{+3}$/H$^+$& \nodata    &    \nodata    &\nodata    & \nodata&  4.21E-08& \nodata &  7.40E-08&    \nodata&    \nodata    &  5.33E-08  \\
ICF(Cl)& 1.00&      1.00&      1.01&      1.01&      1.54&      1.00&      1.10&      1.02&      1.79&      1.01     \\
Ar$^{+2}$/H$^+$& 5.16E-07&  1.11E-06&  6.54E-07&  1.51E-06&  1.99E-06&  1.35E-06&  1.56E-06&  3.13E-06&  8.23E-07&  1.30E-06  \\
Ar$^{+3}$/H$^+$& \nodata    & \nodata  & \nodata & \nodata & \nodata & \nodata & \nodata & \nodata  & \nodata & \nodata  \\
ICF(Ar)& \nodata&  \nodata&      1.06&      1.22&      1.77&      1.07&      1.12&  \nodata&      1.98&      1.07 \\
T$_{O3}$(K)& 11700&     8500&    10400&     8100&    10500&     8900&     9500&     9100&    11900&     9400  \\
T$_{N2}$(K)& 11400&    10700&     6700&     7300&     9400&     8700&    10200&     8500&    11400&    10200  \\
T$_{O2}$(K)& 10100&    17100&     8800&     6800&     5800&     6700&    10600&     8600&     6800&    10100  \\
T$_{S2}$(K)& 7900&     8700&  \nodata&   (101000)&    12900&   (117400) &    13900&    13600&        \nodata&     9200 \\
T$_{S3}$(K)& 12700&     8100&    10300&     9000&    11400&     9900&    10600&     5100&    10400&     9500 \\
N$_{e,S2}$(cm$^{-3}$)& 4000&     3300&      800&     1700&      600&     1700&     4300&       100&     2000&     4100 \\

\enddata

\tablenotetext{a}{Line strengths from which these abundances were derived can be found in Kwitter \& Henry (1998)}
\tablenotetext{b}{Line strengths from which these abundances were derived can be found in Kwitter, Henry, \& Bates (2000)}
\tablenotetext{c}{Line strengths from which these abundances were derived can be found in Henry, Kwitter, \& Dufour (1999)}
\tablenotetext{d}{Values refer to position D}
\tablenotetext{e}{Values refer to position B}


\end{deluxetable}

\clearpage

\begin{deluxetable}{lcccccccccccc}
\tabletypesize{\scriptsize}
\setlength{\tabcolsep}{0.03in}
\tablecolumns{13}
\tablewidth{0in}
\tablenum{8B}
\tablecaption{Elemental Abundances}
\tablehead{
\colhead{Element} & 
\colhead{DdDM1}   &
\colhead{IC418}   &
\colhead{IC3568} &
\colhead{IC4593} &
\colhead{NGC 6720} &
\colhead{NGC 6826}   &
\colhead{NGC 7009}   &
\colhead{NGC 7293}   &
\colhead{NGC 2392}   &
\colhead{NGC 6210}   &
\colhead{Sun\tablenotemark{a}} &
\colhead{Orion\tablenotemark{b}}
}
\startdata
He/H&        0.10&      0.07&      0.12&      0.10&      0.12&      0.11&      0.12&      0.13&      0.08&      0.11&      0.10&      0.10     \\
O/H ($\times 10^4$)& 1.38&      1.39&      3.77&      4.98&      7.64&      4.16&      5.63&      6.47&      3.75&      5.21&      4.90&      5.25      \\
N/H ($\times 10^4$)&  0.29&      0.82&      0.18&      0.28&      2.67&      0.50&      2.36&      2.32&      1.14&      1.15&      0.83&      0.60     \\
Ne/H ($\times 10^4$)& 0.23&      0.08&      0.72&      0.91&      2.08&      0.84&      1.31&      4.23&      1.07&      1.32&      1.20&      0.78      \\
S/H ($\times 10^5$)& 0.24&      0.74&      0.12&      0.41&      0.48&      0.28&      0.84&      0.55&      0.56&      0.67&      1.58&      1.48      \\
Cl/H ($\times 10^7$)& 0.21&      \nodata&      0.33&      0.91&      2.46&      0.85&      1.81&      \nodata&      \nodata&      1.39&      1.91&      2.14      \\
Ar/H ($\times 10^6$)& 0.52&      1.11&      1.07&      1.84&      4.29&      1.51&      2.64&      3.13&      1.63&      1.75&      2.51&      3.09      \\
N/O&  0.21&      0.59&      0.05&      0.06&      0.35&      0.12&      0.42&      0.36&      0.30&      0.22&      0.16&      0.11      \\
Ne/O&   0.17&      0.05&      0.19&      0.18&      0.27&      0.20&      0.23&      0.65&      0.28&      0.25&      0.25&      0.15       \\
S/O ($\times 10^1$)& 0.18&      0.53&      0.03&      0.08&      0.06&      0.07&      0.15&      0.09&      0.15&      0.13&      0.32&      0.28     \\
Cl/O ($\times 10^3$)&  0.15&      \nodata&      0.09&      0.18&      0.32&      0.20&      0.32&      \nodata&      \nodata&      0.27&      0.39&      0.41         \\
Ar/O ($\times 10^2$)&  0.37&      0.80&      0.28&      0.37&      0.56&      0.36&      0.47&      0.48&      0.43&      0.34&      0.51&      0.59     \\

\enddata

\tablenotetext{a}{Grevesse \& Sauval (1998; S, Cl, meteoritic; Ne, Ar, photospheric); O abundance from Allende Priete, Lambert, \& Asplund 2001}
\tablenotetext{b}{Esteban et al. (1998), Table 19, gas + dust}

\end{deluxetable}

\begin{figure}[h]
\centering
\figurenum{1}
\includegraphics[width=12cm,angle=270]{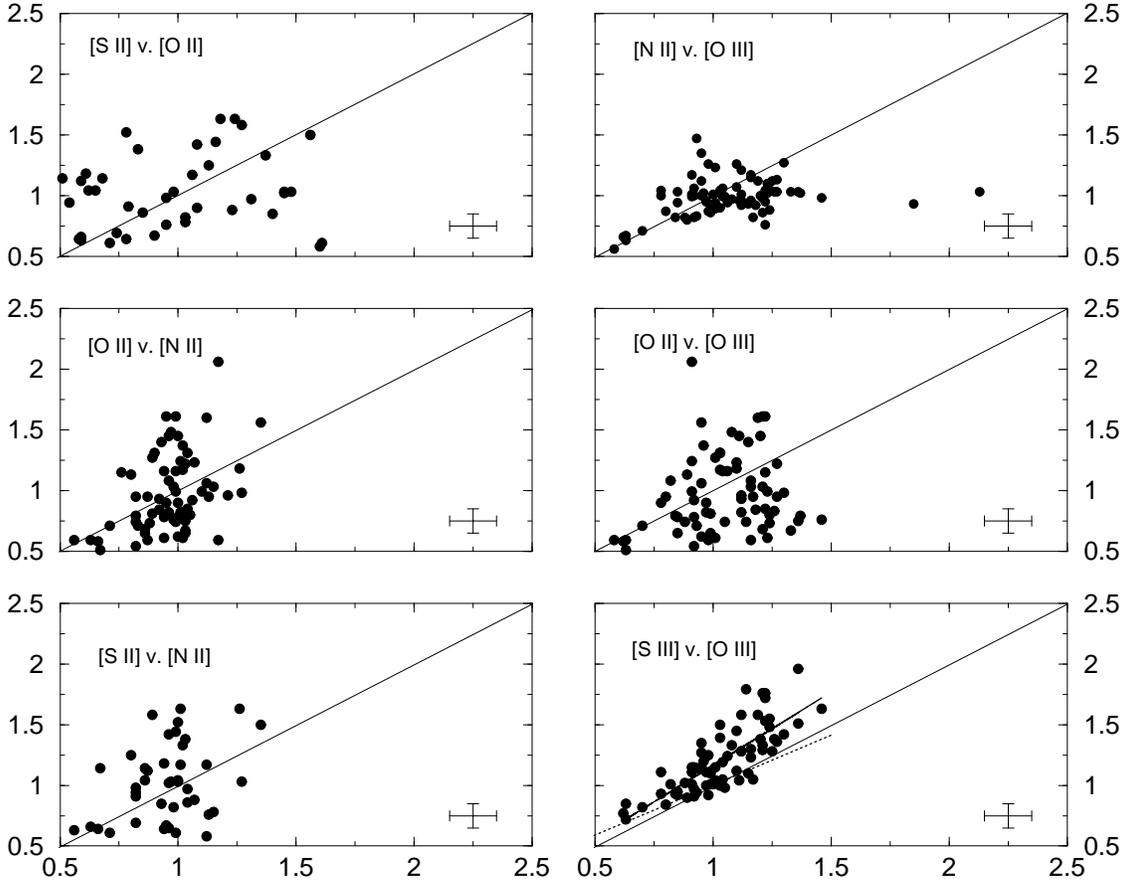}
\caption{Comparisons of five electron temperatures (in units of 10$^4$K) for our sample objects in various combinations. Each panel is labeled to show the temperatures being plotted, Y v. X. The diagonal lines show the one-to-one correspondence. In the lower right panel, the solid bold line shows our least squares fit to the data (eq.~1), while the dashed line shows the functional form derived by Garnett (1992) in a photoionization model study of H~II regions.}
\end{figure}

\clearpage
\begin{figure}[h]
\centering
\figurenum{2}
\includegraphics[width=12cm,angle=270]{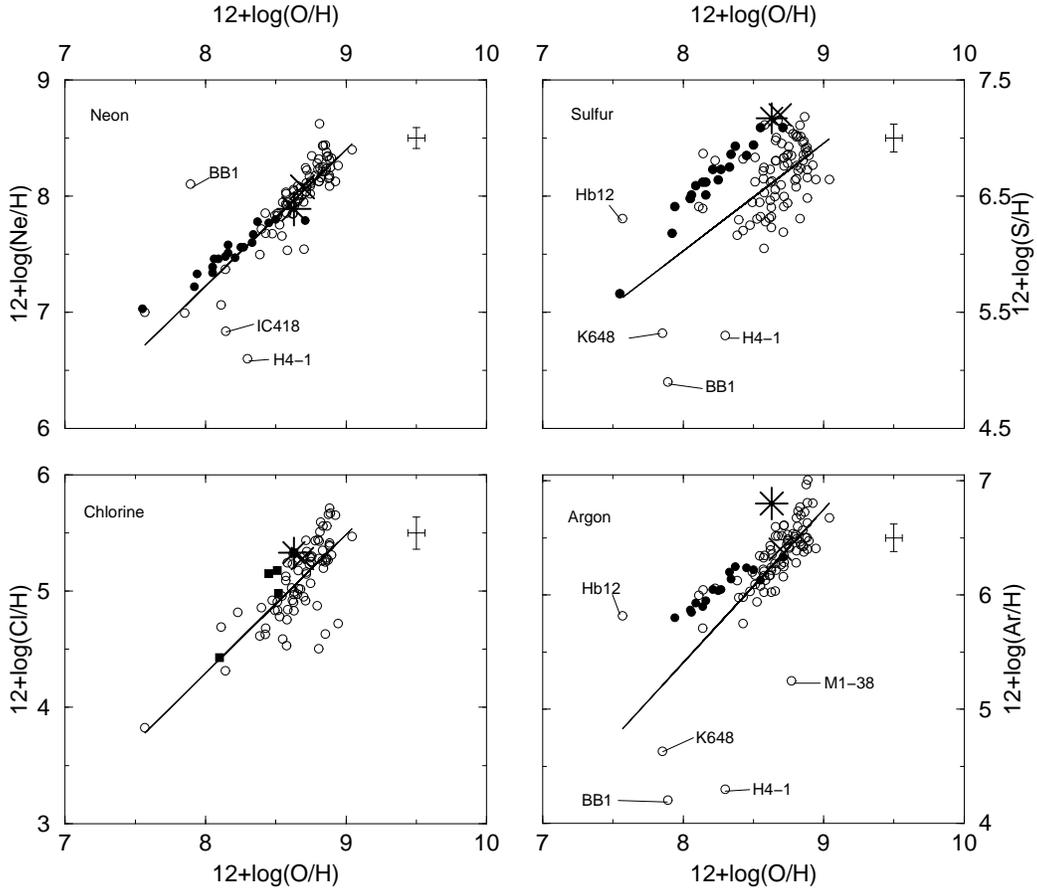}
\caption{Plots of 12+log(X/H) versus 12+log(O/H), where X is the element specified in the upper left of each graph. Results are shown for both the PNe in the current sample (open circles) as well as the data for H~II regions in M101 as measured by KBG (filled circles), Orion (Esteban et al. 1998; star) and the Sun (Grevesse \& Sauval 1998 but with oxygen abundance from Allende Prieto et al. 2001; X). H~II region data for Cl are taken from references in Table~3 and are shown as filled squares. Solid, bold lines show least squares fits to our PN data, where the slopes and intercepts for the regressions are provided in Table~4. A representative error bar is shown in each panel. Certain outliers are identified in the figure and discussed in the text.}
\end{figure}

\clearpage

\begin{figure}[h]
\centering
\figurenum{3a}
\includegraphics[width=12cm,angle=270]{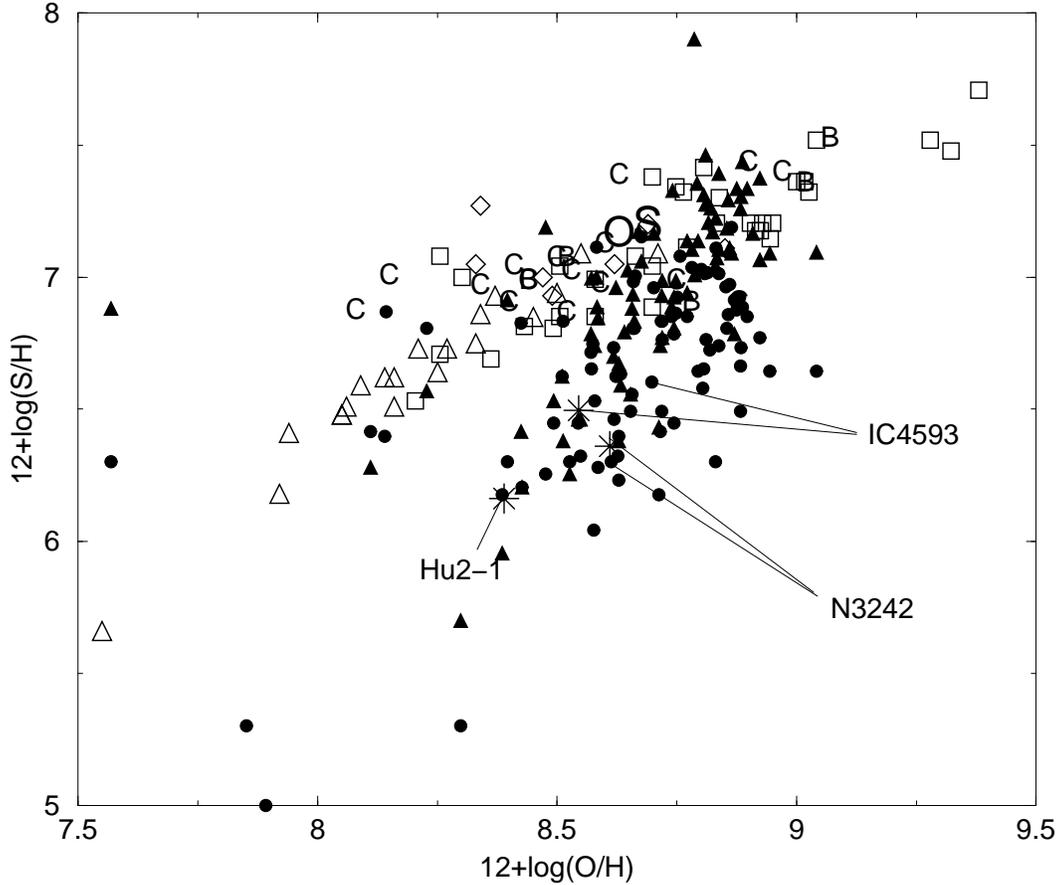}
\caption{12+log(S/H) versus 12+(O/H) for PNe from this paper (filled symbols) and H~II regions from several sources (open symbols).  Filled circles show PN results in which S$^{+2}$ results for our sample were derived from the [S~III] $\lambda\lambda$9069,9532 lines, while filled triangles show analogous results using the [S~III] 6312{\AA} line. H~II region abundances are taken from Shaver et al. (1983; open diamonds), Afflerbach, Churchwell, \& Werner 1997; open squares), and KBG (open triangles). Also shown are results for the Orion Nebula from Esteban et al. (1998; O), the Sun (Grevesse \& Sauval 1998 for the S abundance, Allende Prieto et al. 2001 for the O abundance; S), Cepheids from Andrievsky et al. (2002a,b,c; C) and for B giants (Trundle et al. 2002; B). Model-predicted values for Hu2-1, IC4593, and NGC~3242 are shown with stars (see text for explanation). Uncertainties for each data set are [in the form $\pm$dlog(S/H),$\pm$dlog(O/H)]: Shaver et al., 0.06,0.04; Afflerbach et al., 0.22,0.35; KBG, 0.11,0.07; this paper, 0.12,0.06.}
\end{figure}

\clearpage

\begin{figure}[h]
\centering
\figurenum{3b}
\includegraphics[width=12cm,angle=270]{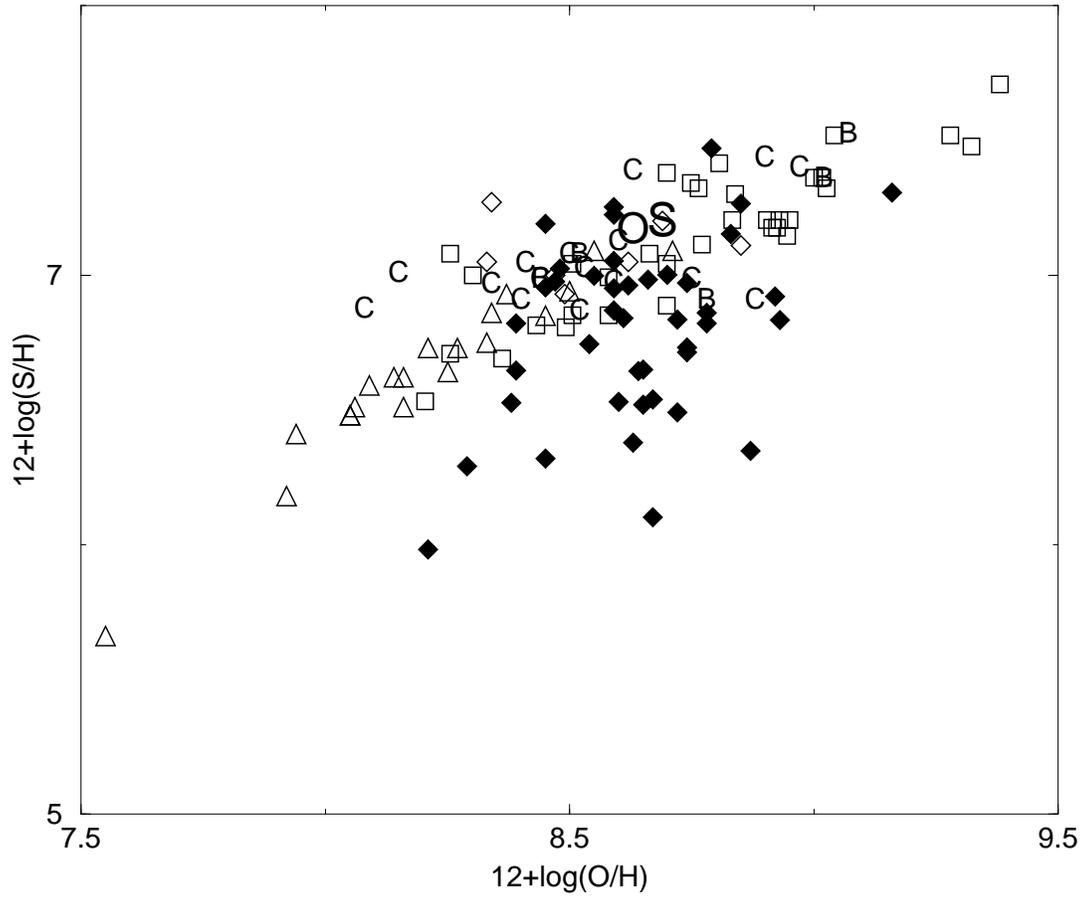}
\caption{Same as 3a but showing the KB PN sample as filled diamonds.}
\end{figure}

\clearpage

\begin{figure}[h]
\centering
\figurenum{4}
\includegraphics[width=10cm,angle=270]{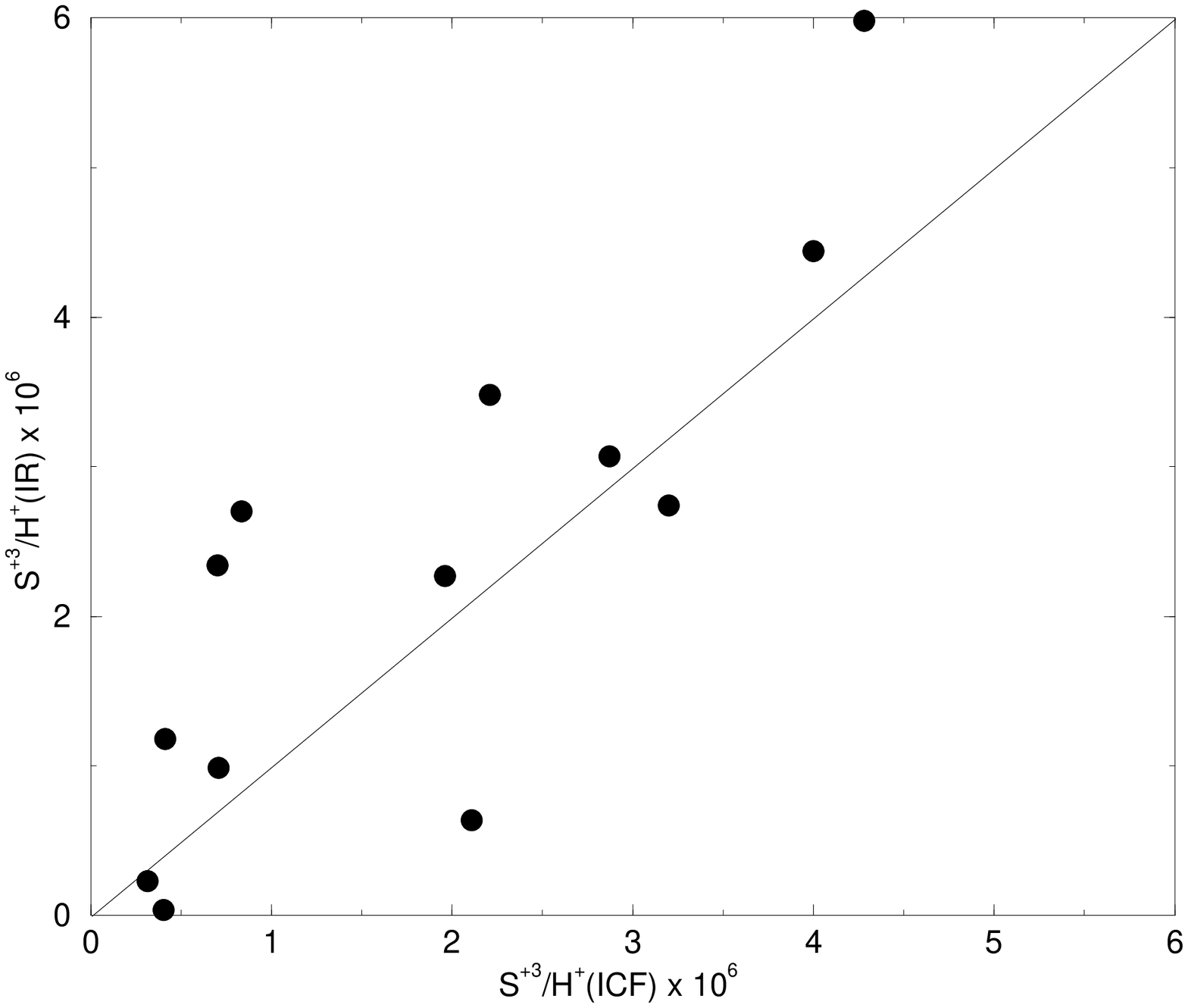}
\caption{S$^{+3}$/H$^+$ determined from direct IR measurments of [S~IV] 10.5$\mu$m versus the same parameter inferred from our ICF. Both sets of numbers are taken directly from columns~3 and 2 in Table~6, respectively. The diagonal line shows the one-to-one correspondence.}
\end{figure}

\clearpage

\begin{figure}[h]
\centering
\figurenum{5}
\includegraphics[width=10cm,angle=270]{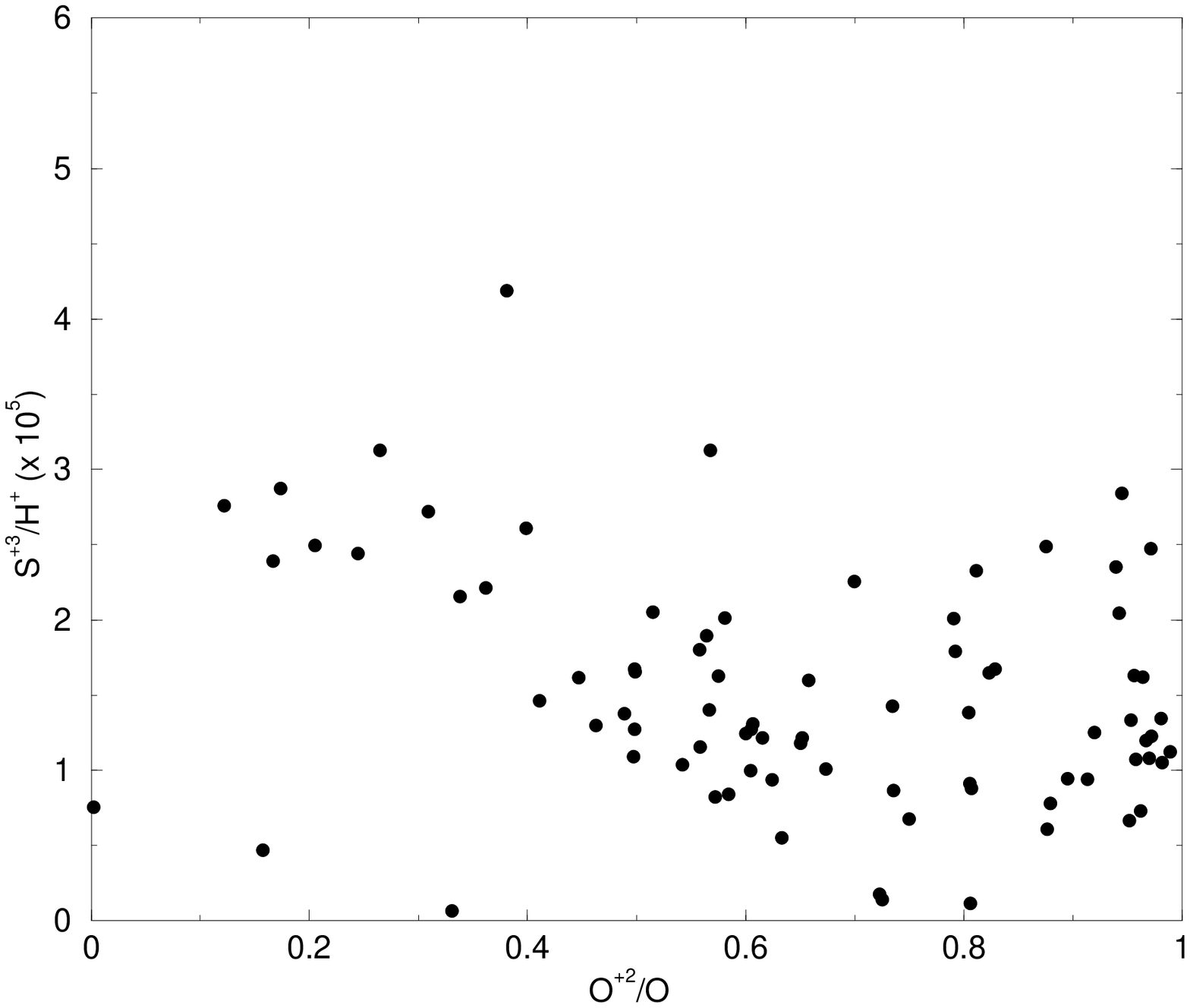}
\caption{Predicted values of the sulfur deficit (see text) in units of S$^{+3}$/H$^+$ (x 10$^5$) versus observed results for O$^{+2}$/O for members of our PN sample.}
\end{figure}

\clearpage

\begin{figure}
\figurenum{6a}
\centering
\leavevmode
\includegraphics[width=12cm,angle=270]{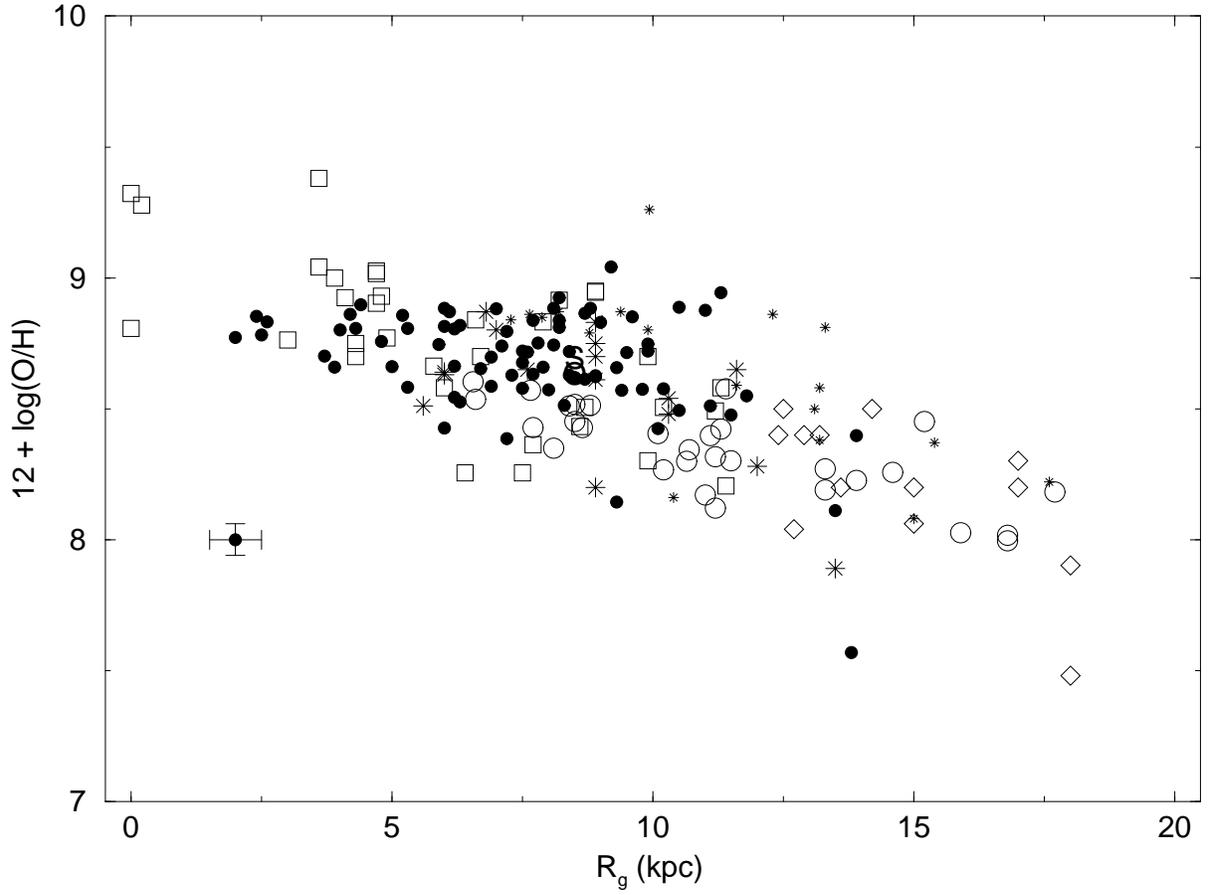}
\caption{12+log(O/H) versus galactocentric distance in kiloparsections. Our abundance results are shown with filled circles. H~II region abundances as determined by Afflerbach, Churchwell, \& Werner (1997; open boxes), Deharveng et al. (2000; open circles) and V{\'i}lchez \& Esteban (1996; open diamonds) are shown for comparison, as are B main sequence stellar data from Gummersbach et al. (1998; large stars) and Rolleston et al. (2000; small stars).}
\end{figure}

\begin{figure}
\figurenum{6b}
\centering
\leavevmode
\includegraphics[width=12cm,angle=270]{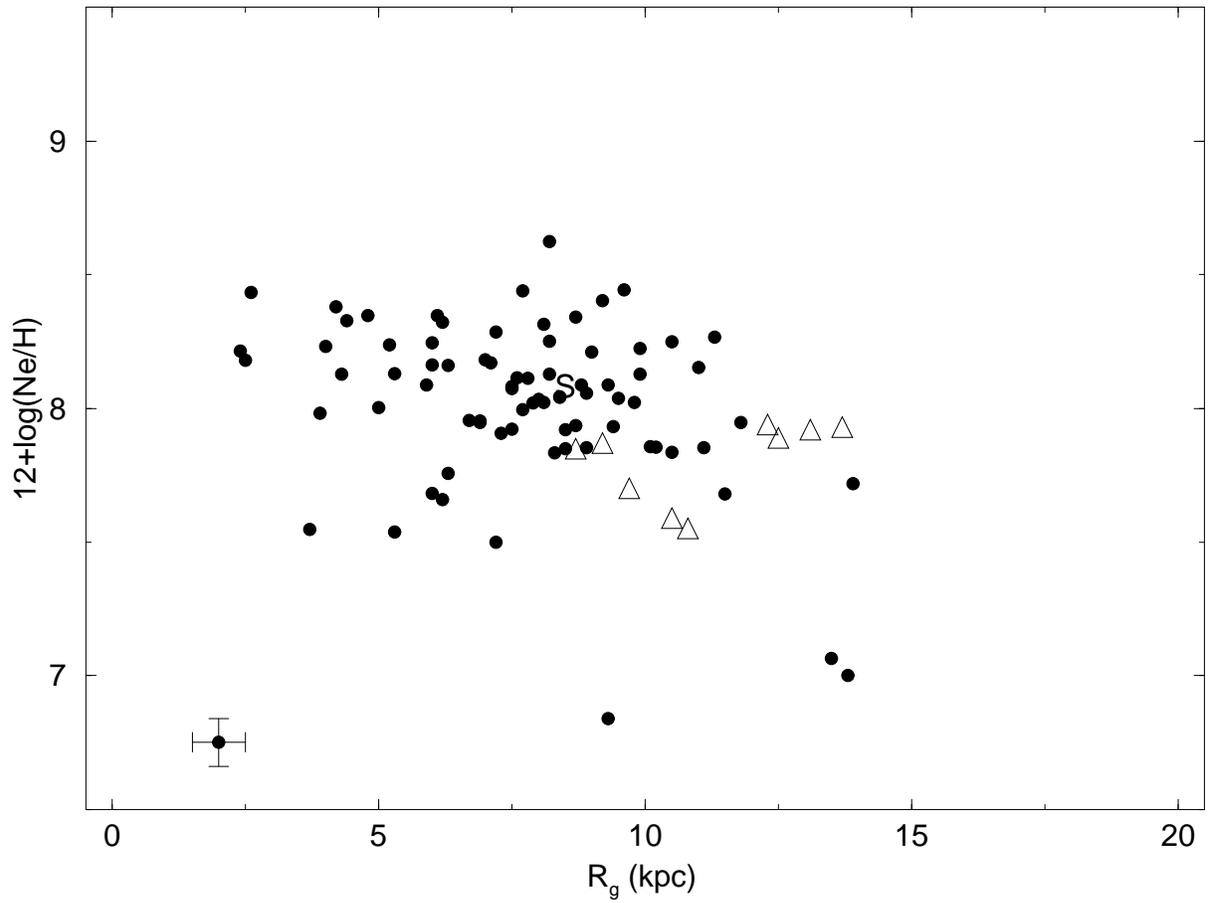}
\caption{Same as Fig.~6a but for neon. Data from Shaver et al. (1983; up triangles) are shown for comparison.}
\end{figure}

\begin{figure}
\figurenum{6c}
\centering
\leavevmode
\includegraphics[width=12cm,angle=270]{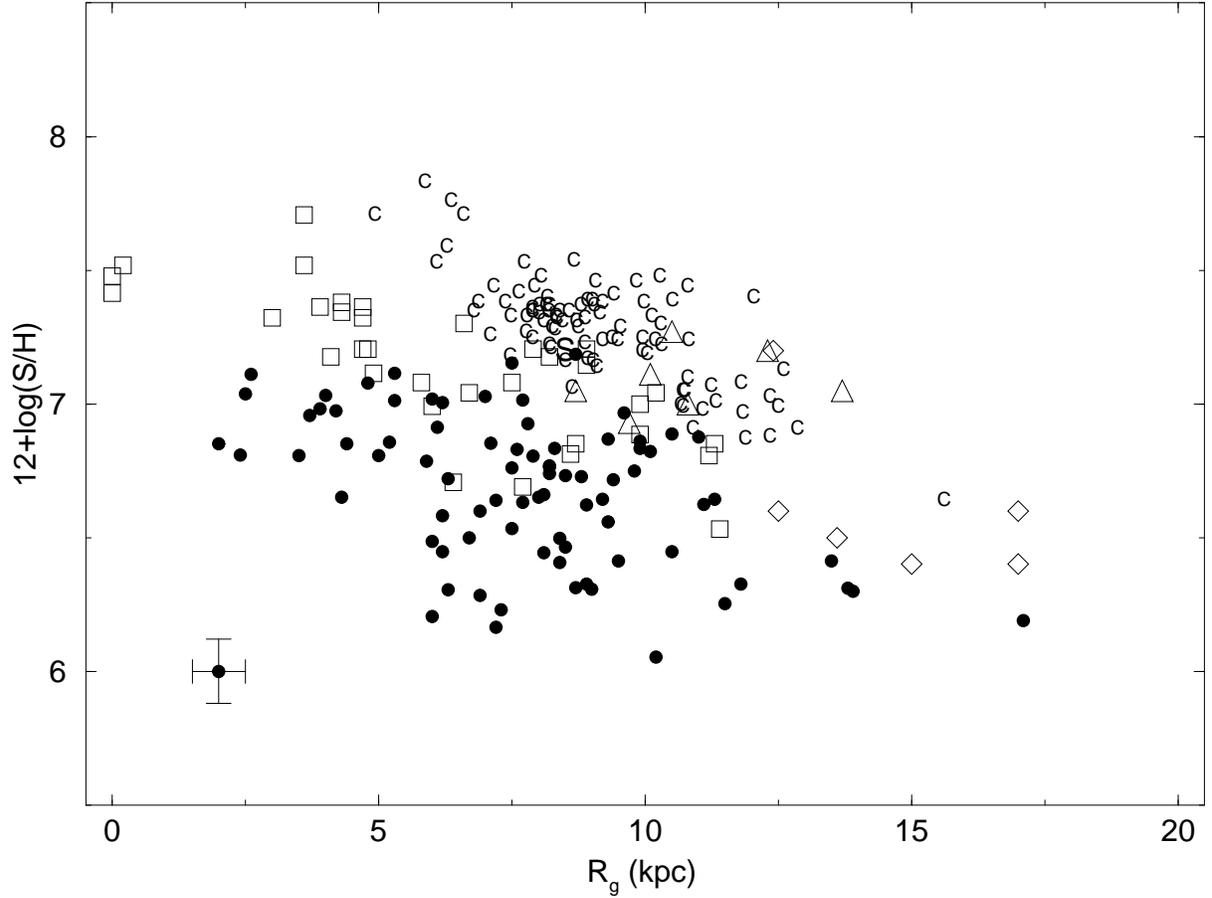}
\caption{Same as Fig.~6a but for sulfur. Data from Shaver et al. (1983; up triangles), Afflerbach, Churchwell, \& Werner (1997; open boxes), and  V{\'i}lchez \& Esteban (1996; open diamonds) for H~II regions as well as data for 100 Cepheids from Andrievsky et al. (2002a,b,c; C) are shown for comparison.}
\end{figure}

\begin{figure}
\figurenum{6d}
\centering
\leavevmode
\includegraphics[width=12cm,angle=270]{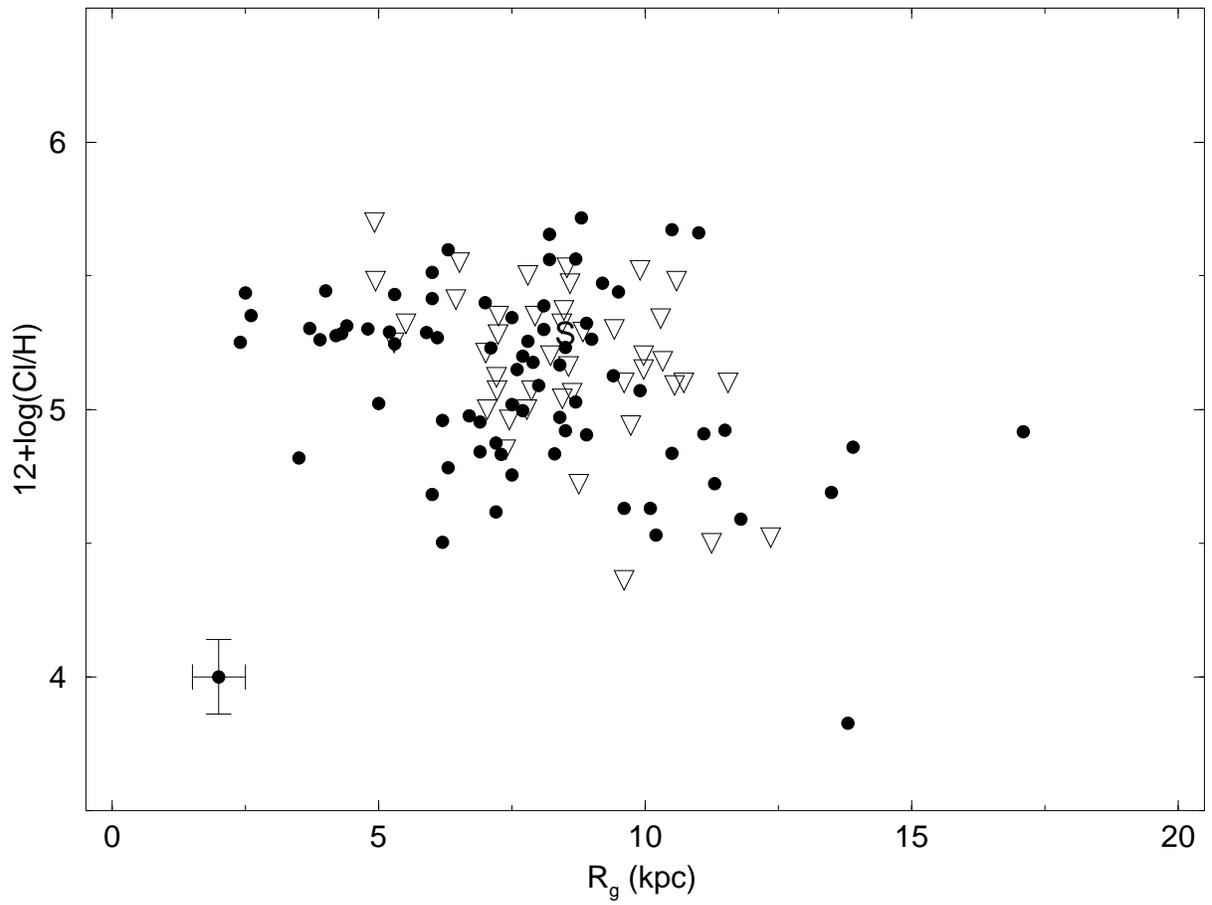}
\caption{Same as Fig.~6a but for chlorine. PN data from Maciel \& Chiappini (1994; down triangles) are shown for comparison.}
\end{figure}

\begin{figure}
\figurenum{6e}
\centering
\leavevmode
\includegraphics[width=12cm,angle=270]{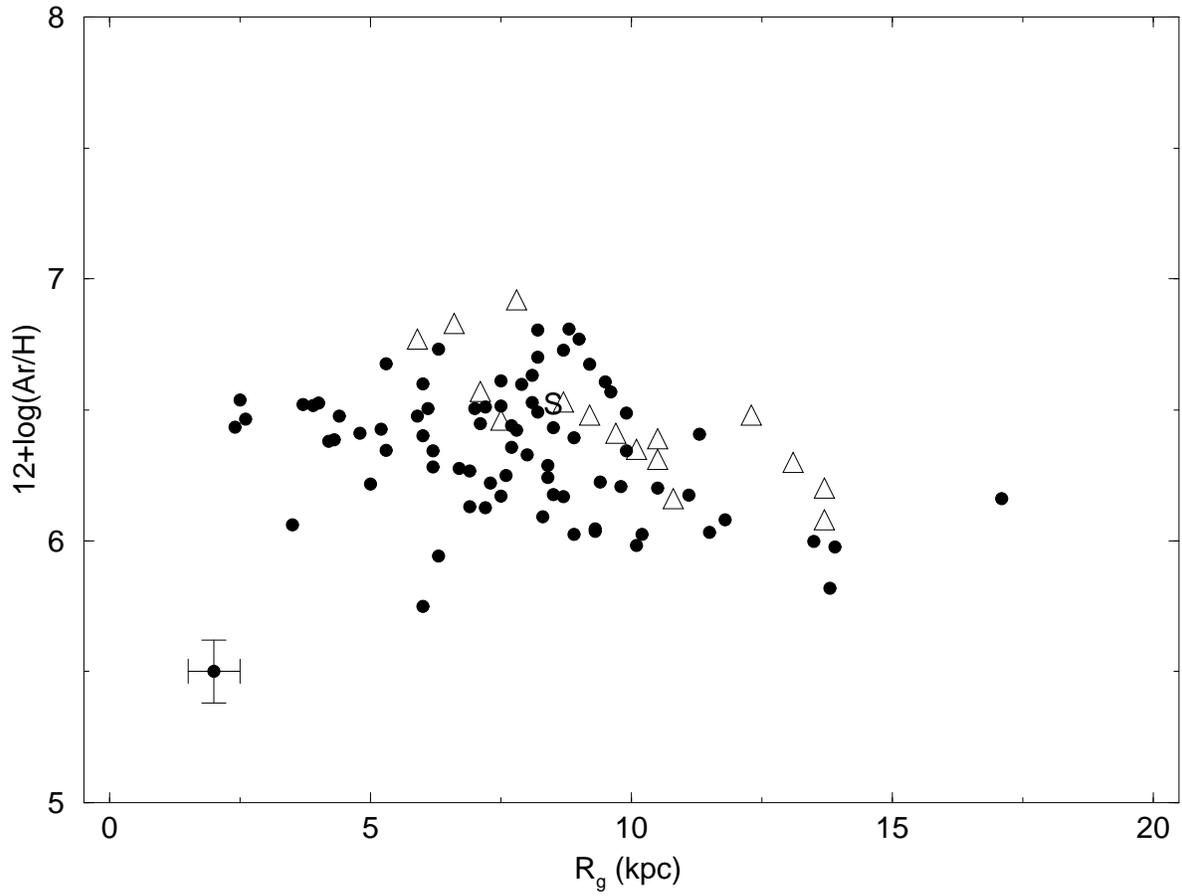}
\caption{Same as Fig.~6a but for argon. Data from Shaver et al. (1983; up triangles) for H~II regions are shown for comparison.}
\end{figure}

\end{document}